\documentclass[extra, referee]{gji}

\usepackage{timet}
\usepackage{url} 
\usepackage{lineno}
\usepackage{soul}
\usepackage{nccmath}
\usepackage{enumitem}
\usepackage{ifthen}
\usepackage{color}
\usepackage{graphicx}
\usepackage{amsmath}
\usepackage{amssymb}
\usepackage{algorithm}
\usepackage{algorithmic}
\usepackage{setspace}
\usepackage{multirow}
\usepackage[T1]{fontenc}

\DeclareMathOperator*{\argmax}{argmax}
\DeclareMathOperator*{\argmin}{argmin}
\graphicspath{{Fig/}}
\title[Variational Prior Replacement]{Variational Prior Replacement in Bayesian Inference and Inversion}
\author[X. Zhao and A. Curtis]
{Xuebin Zhao$^1$ and Andrew Curtis$^{1}$\\
	$^1$ School of GeoSciences, University of Edinburgh,
	Edinburgh, United Kingdom \\
}

\begin{document}
	\label{firstpage}
	
	
	\begin{titlepage}
		\begin{center}
			\vspace*{2.0cm} 
			
			\huge
			\textbf{Variational Prior Replacement in Bayesian Inference and Inversion}
			
			\vspace{1.0cm}
			\LARGE
			Xuebin Zhao$^1$ and Andrew Curtis$^{1}$\\
			
			\vspace{0.5cm}
			\Large
			$^1$ School of GeoSciences, University of Edinburgh, United Kingdom \\
			
			\vspace{0.4cm}
			\Large
			E-mail: \textit{xuebin.zhao@ed.ac.uk, andrew.curtis@ed.ac.uk}
			
			\vfill
			\vfill
		\end{center}
	\end{titlepage}
	
	\newpage

\begin{summary}
Many scientific investigations require that the values of a set of model parameters are estimated using recorded data. In Bayesian inference, information from both observed data and prior knowledge is combined to update model parameters probabilistically by calculating the posterior probability distribution function. Prior information is often described by a prior probability distribution. Situations arise in which we wish to change prior information during the course of a scientific project. However, estimating the solution to any single Bayesian inference problem is often computationally costly, as it typically requires many model samples to be drawn, and the data set that would have been recorded if each sample was true must be simulated. Recalculating the Bayesian inference solution every time prior information changes can therefore be extremely expensive. We develop a mathematical formulation that allows the prior information that is embedded within a solution, to be changed using variational methods, without recalculating the original Bayesian inference. In this method, existing prior information is removed from a previously obtained posterior distribution and is replaced by new prior information. We therefore call the methodology \textit{variational prior replacement} (VPR). We demonstrate VPR using a 2D seismic full waveform inversion example, in which VPR provides similar posterior solutions to those obtained by solving independent inference problems using different prior distributions. The former can be completed within minutes on a laptop computer, whereas the latter requires days of computations using high-performance computing resources. We demonstrate the value of the method by comparing the posterior solutions obtained using three different types of prior information: uniform, smoothing and geological prior distributions.
\end{summary}

\begin{keywords}
Bayesian Inference, Probability Distributions, Seismic Tomography, Inverse Theory, Waveform Inversion
\end{keywords}

\newpage

\section{Introduction}
In a wide variety of scientific and engineering applications, researchers seek to estimate unknown (latent) parameters using observed data by solving an inverse or inference problem. By approximating the physical system, it is usually possible to calculate a forward function which estimates the synthetic data that would have been observed if any particular set of model parameter values were true, and this function commonly has unique values. However, direct inversion of this function is difficult if not impossible, due to uncertainties in the finite number of measurable data, and to the nonlinearity of the forward function \citep{boyd2004convex}. Typically in practise, solutions to such inverse problems are non-unique \citep{mosegaard1995monte, mosegaard2002monte, tarantola2005inverse, valentine2023emerging}. 

Bayesian inference solves fully nonlinear, non-unique inverse problems under a probabilistic framework by seeking to define the family of all plausible solutions within model parameter space. The solution to Bayesian inference is described by the so-called posterior probability distribution function (\textit{pdf} - either a probability density function for continuous variables, or a set of probabilities for discrete variables), obtained by updating prior information about model parameters with new information from observed data. It provides a statistical description of how consistent is each solution with both the data and prior information, and allows uncertainties in the inverse problem solution to be estimated \citep{tarantola2005inverse, arnold2018interrogation}. 

Global search or sampling based methods are often used to solve Bayesian inference problems. Samples of parameter values with non-zero posterior probability values are retained, sometimes in proportion to their probabilities, to build an ensemble of model solutions. These solutions are used to estimate statistical properties of the posterior distribution that characterise the solution uncertainty. Monte Carlo is one of the most frequently used sampling methods, including Metropolis-Hastings Markov chain Monte Carlo \cite[MH-McMC --][]{metropolis1953equation, hastings1970monte, press1968earth, mosegaard1995monte, sambridge2002monte}, trans-dimensional Monte Carlo \citep{green1995reversible, malinverno2002parsimonious, sambridge2006trans, bodin2009seismic, galetti2017transdimensional}, gradient-based Monte Carlo methods \citep{welling2011bayesian, girolami2011riemann, fichtner2019hamiltonian, gebraad2020bayesian, zhao2021gradient, biswas2022transdimensional, de2023acoustic, berti2023computationally} and informed-proposal Monte Carlo \citep{khoshkholgh2021informed, khoshkholgh2022full}. Global search methods that do not use Markov chains have been developed to solve Bayesian problems, either using optimisation to search for the most probable solution such as in simulated annealing \citep{kirkpatrick1983optimization, sen2013global, zhao2022hybrid} and genetic algorithms \citep{stoffa1991nonlinear, sambridge1992genetic}, or algorithms that characterise the posterior distribution such as the neighbourhood algorithm \citep{sambridge1999geophysical, sambridge1999geophysical2}, prior sampling \citep{meier2007fully, meier2007global, kaufl2016solving, mosser2020stochastic, bloem2022introducing}, exact sampling \citep{propp1996exact, walker2014spatial} and direct estimation of posterior pdfs without sampling \citep{nawaz2016bayesian}. However, all such sampling methods present deficiencies when applied to complex or high dimensional inference problems such as slow convergence \citep{atchade2005adaptive, andrieu2008tutorial}, which implies that a large number of model samples and their corresponding forward simulations are required.

Variational inference provides an alternative to random sampling methods to solve Bayesian problems. Variational methods select one optimal approximation to the true posterior pdf from a predefined family of known and computationally tractable probability distributions (referred to as the variational family). This is accomplished by minimising the difference between the posterior and variational pdfs \citep{bishop2006pattern, blei2017variational, zhang2021an}, thus solving Bayesian problems using optimisation rather than stochastic sampling. This approach can often be computationally efficient, ease the detection of convergence, and scale well to high dimensional inference problems with large datasets, while still producing a valid probability distribution. 

In geophysics, variational inference was first applied to estimate subsurface geological facies and petrophysical parameters using seismic data \citep{nawaz2018variational, nawaz2019rapid, nawaz2020variational}, where a mean-field approximation (which ignores correlations between parameters) is employed to simplify the mathematical formulation of the variational problem \citep{bishop2006pattern, kucukelbir2017automatic}. Since then, more advanced variational methods have been developed for different geophysical problems, such as travel time tomography \citep{zhang2019seismic, zhao2021bayesian, levy2022variational}, seismic migration \citep{siahkoohi2021preconditioned, siahkoohi2023reliable}, seismic amplitude inversion \citep{zidan2022regularized}, earthquake hypocentre inversion \citep{smith2022hyposvi}, slip distribution inversion \citep{sun2023new}, full waveform inversion \citep{zhang2021bayesianfwi, bates2022probabilistic, wang2023re, lomas20233d, izzatullah2023physics, zhao2024physically, yin2024wiser}, and experimental design \citep{strutz2023variational}.

Most studies mentioned above, whether using random sampling or variational methods, focus on performing Bayesian inference efficiently and accurately given a specific set of observed data and fixed prior knowledge. Over recent years, researchers made use of neural networks and other machine learning architectures to implement efficient Bayesian inference in which the posterior pdf can be obtained rapidly for any newly observed dataset \citep{devilee1999efficient, meier2007fully, meier2007global, shahraeeni2011fast, shahraeeni2012fast, de2013bayesian, kaufl2014framework, kaufl2016solving, earp2019probabilistic, earp2019probabilistic2, zhang2021bayesian, mosher2021probabilistic, wang2022gaussian, hansen2022use, alyaev2022direct, grana2022probabilistic, guan2023efficient, sun2024invertible}. However, almost no publications consider situations where we want to change (update) the prior information used in a previously performed inference process, or where we have multiple plausible prior hypotheses to be tested for the same observed data. In such cases one might have to perform the inference repeatedly with different prior distributions; this would become extremely expensive in many applications, even though the data used in each individual case do not change. 

\cite{walker2014varying} introduced a method called prior replacement, which allows prior information to be changed rapidly in Bayesian inference without repeating the full inference procedure for each individual prior pdf (on occasion below, we may refer to this simply as the prior). This is achieved by dividing the obtained posterior distribution by the current prior pdf in an attempt to remove the effect of the latter, and then multiplying the result by a new prior pdf to inject (update) the results with new prior information. The method was demonstrated to be effective for varying prior information when estimating rock physics parameters (clay volume and sandstone matrix porosity) using seismic impedance data. \cite{walker2014varying} used (semi-)analytic methods to perform prior replacement, which requires the calculation of integrals of probability distributions over the entire parameter space, making it difficult to calculate for high dimensional problems. Moreover, the analytic calculation is only applicable under stringent conditions: first, the existing posterior distribution is represented by a mixture of Gaussian distributions. And second, the old and new prior distributions should be uniform, Gaussian, or (possibly) other probability distributions whose integral over the parameter space and whose multiplication and division by Gaussian distributions are analytically tractable, such that the replacement of the new posterior distribution can be calculated analytically.

In this paper, we develop a prior replacement methodology under the framework of variational inference, hence the name \textit{variational prior replacement} (VPR). VPR addresses (relaxes) the issues mentioned above, making it applicable for high dimensional and complicated Bayesian inference problems. We test and demonstrate the method on a full waveform inversion (FWI) problem. Until the last few years there was no published fully nonlinear Bayesian solution to any FWI problem that approached a practical scale, due to the huge computational cost, and associated theoretical and algorithmic challenges \citep{gebraad2020bayesian, zhang2020variational}. While to-date a single study has extended the method to a three dimensional case \citep{zhang20233}, the computational issues to make this approach mainstream remain. Advances that deliver even approximate results using greatly reduced computation are therefore significant.

The rest of this paper is organised as follows. In Section 2, we review the Bayesian inference and prior replacement concept developed in \cite{walker2014varying}. To perform prior replacement more efficiently, we introduce variational inference and derive the variational prior replacement (VPR) framework. In Section 3, we demonstrate the method using a seismic full waveform inversion example, in which we compare the results obtained using VPR with those found using independent Bayesian inference for each prior pdf. We demonstrate the effectiveness of the method by testing three different prior distributions using the same observed data. Finally, we provide a brief discussion and draw conclusions from this study.

\section{Methodology}
\subsection{Bayesian Inference}
In Bayesian inference, inverse problems are solved under a probabilistic framework by calculating the so-called \textit{posterior} probability distribution function (pdf) of model vector $\mathbf{m}$ given observed data $\mathbf{d}_{obs}$ using Bayes' rule:
\begin{equation}
p(\mathbf{m}|\mathbf{d}_{obs}) = \dfrac{p(\mathbf{d}_{obs}|\mathbf{m})p(\mathbf{m})}{p(\mathbf{d}_{obs})}
\label{eq:bayes}
\end{equation}
where $p(\mathbf{m})$ is the \textit{prior} pdf that describes available information about model parameter $\mathbf{m}$ before inference process, and $p(\mathbf{d}_{obs}|\mathbf{m})$ is the \textit{likelihood} function, which calculates the probability of observing data $\mathbf{d}_{obs}$ given any model value $\mathbf{m}$. The likelihood is used to describe how well $\mathbf{d}_{obs}$ matches synthetic data generated by a particular model $\mathbf{m}$. Term 
\begin{equation}
p(\mathbf{d}_{obs})=\int_{\mathbf{m}}p(\mathbf{d}_{obs}|\mathbf{m})p(\mathbf{m})d\mathbf{m}
\label{eq:evidence}
\end{equation}
is a normalisation constant called the \textit{evidence}. It ensures that the right hand side of equation \ref{eq:bayes} is a valid probability distribution. Bayesian inference combines information from both data and prior knowledge in a probabilistic manner, and the resulting posterior distribution describes all possible model solutions that fit the data and the prior.

\subsection{Prior Replacement}
Consider a situation where we have different sets of prior information (e.g., different hypotheses, or differing beliefs held by different people) about model parameter $\mathbf{m}$, defined by prior probability distributions $p_1(\mathbf{m})$, $p_2(\mathbf{m})$, and so on, and we wish to evaluate the implications of these various priors by calculating the corresponding posterior distributions. Such an array of prior distributions might originate from the views of different groups of experts, or might invoke different assumptions about the structures and properties that might pertain to the model, perhaps representing a range of different hypotheses to be tested and discriminated \cite[e.g.,][]{bloem2022introducing}. A straightforward strategy is to apply Bayes' rule to each prior distribution, and solve independent (prior specific) Bayesian inverse problem whenever prior information changes. However, such a prior specific approach is not practically feasible, since it is already expensive to perform a single inference process, especially for high dimensional problems with large datasets such as typically occurs in seismic tomography and FWI problems.

Alternatively, suppose we have obtained a posterior distribution for data $\mathbf{d}_{obs}$ based on one specific type of prior information. When additional prior information becomes available or when different prior hypotheses exist and need to be discriminated, we might remove the effect of the existing prior information from the current posterior distribution, then inject different prior information. Thus we would obtain the desired posterior pdf without explicitly applying Bayes' rule a second time. Below we denote prior and posterior pdfs that have been considered or calculated previously as \textit{old} pdfs and those obtained by updating the prior distribution in this way as \textit{new} ones; the words \textit{old} and \textit{new} in this context do not refer to situations where we update information because additional data have been collected, but rather to the order in which different prior distributions are combined with information in a fixed data set.

\cite{walker2014varying} mathematically formulated the above idea as follows: the new posterior distribution $p_{new}(\mathbf{m}|\mathbf{d}_{obs})$ given the new prior distribution $p_{new}(\mathbf{m})$ can be calculated by
\begin{equation}
p_{new}(\mathbf{m}|\mathbf{d}_{obs}) = \dfrac{p(\mathbf{d}_{obs}|\mathbf{m})p_{new}(\mathbf{m})}{p_{new}(\mathbf{d}_{obs})} = \dfrac{p(\mathbf{d}_{obs}|\mathbf{m})p_{old}(\mathbf{m})}{p_{old}(\mathbf{d}_{obs})}\ \dfrac{p_{new}(\mathbf{m})}{p_{old}(\mathbf{m})}\ \dfrac{p_{old}(\mathbf{d}_{obs})}{p_{new}(\mathbf{d}_{obs})}
\label{eq:bayes_new_from_old}
\end{equation}
The first equality is simply Bayes' rule. In both new and old distributions, we assume that the observed data are the same. On the right hand side, $p_{old}(\mathbf{d}_{obs})$ and $p_{new}(\mathbf{d}_{obs})$ are two constants which are independent of model vector $\mathbf{m}$ according to equation \ref{eq:evidence}, so we define $k = p_{old}(\mathbf{d}_{obs}) / p_{new}(\mathbf{d}_{obs})$ for later convenience. Term $p_{new}(\mathbf{m})/p_{old}(\mathbf{m})$ essentially takes the role of changing (replacing) the old prior by the new prior pdf in Bayesian inference, and states how we inject new prior information. Denote
\begin{equation}
p_{old}(\mathbf{m}|\mathbf{d}_{obs}) = \dfrac{p(\mathbf{d}_{obs}|\mathbf{m})p_{old}(\mathbf{m})}{p_{old}(\mathbf{d}_{obs})}
\label{eq:bayes_old}
\end{equation}
as the old posterior distribution given the old prior $p_{old}(\mathbf{m})$. Then equation \ref{eq:bayes_new_from_old} becomes
\begin{equation}
p_{new}(\mathbf{m}|\mathbf{d}_{obs}) = k\ p_{old}(\mathbf{m}|\mathbf{d}_{obs})\ \dfrac{p_{new}(\mathbf{m})}{p_{old}(\mathbf{m})}
\label{eq:bayes_new}
\end{equation}
Equation \ref{eq:bayes_new} has a form that allows us to evaluate the new posterior distribution from the old one by updating (replacing) prior information after Bayesian inference, assuming that we know both $p_{old}(\mathbf{m})$ and $p_{new}(\mathbf{m})$, and that we can evaluate the normalisation constant $k$. Using this formulation, there is no need to perform new likelihood evaluations, which is normally the most computationally expensive step in solving an inverse problem,
no matter how many different prior distributions we wish to inject. However, prior replacement is valid only under one condition: the new prior must have zero (or in practise, very small) probability where the old prior has zero probability values to avoid a numerically unstable situation of dividing by zero \citep{walker2014varying}. Intuitively, the support of the new prior pdf must be a subset of that of the old one.

Equations \ref{eq:bayes_old} and \ref{eq:bayes_new} define the two main operations involved in prior replacement. The former is the solution to a typical Bayesian problem in which the (old) posterior pdf is evaluated given the observed data and existing prior information. The latter can be viewed as a quasi-Bayesian problem in the sense that Bayes rule applied to the new prior distribution is implicit within the formula, and the new posterior pdf is obtained by combining information from three probability distributions -- of similar form to Bayes' rule, but without the need to re-calculate the likelihood function from scratch. Calculation of these expressions can nevertheless be computationally expensive, so in the following two sections we introduce efficient methods for each of these operations, respectively.

\subsection{Variational Inference}
The Bayesian posterior distribution in equation \ref{eq:bayes_old} can be estimated using either random sampling or variational inference methods. Markov chain Monte Carlo (McMC) is a typical sampling method that generates an ensemble of samples distributed according to the posterior distribution as the number of samples tends to infinity \citep{mosegaard1995monte}. However, McMC can be expensive in practise since the required number of samples increases exponentially with the dimensionality of model vector $\mathbf{m}$ -- a concept referred to as the curse of dimensionality \citep{curtis2001prior}. 

Variational inference is an alternative to McMC that can be more efficient in certain situations. In variational inference, we define a family of probability distributions $\mathcal{Q}(\mathbf{m})=\{q(\mathbf{m})\}$ with fixed (predefined) complexity, within which we select a member $q^*(\mathbf{m})$ that best approximates the unknown posterior distribution. Therefore, variational inference solves Bayesian problems using optimisation rather than random sampling. The optimal distribution can be found by minimising the discrepancy between the variational and posterior distributions. 

The Kullback-Leibler (KL) divergence \citep{kullback1951information} is often used to measure the distance between two distributions
\begin{equation}
\text{KL}[q(\mathbf{m})||p(\mathbf{m}|\mathbf{d}_{obs})] = \mathbb{E}_{q(\mathbf{m})}[\log q(\mathbf{m}) - \log p(\mathbf{m}|\mathbf{d}_{obs})]
\label{eq:kl}
\end{equation}
where the expectation is taken with respect to the variational distribution $q(\mathbf{m})$. The KL divergence is non-negative and equals zero only when the two distributions are identical. Substituting Bayes' rule (equation \ref{eq:bayes}) into equation \ref{eq:kl}, we have
\begin{equation}
\log p(\mathbf{d}_{obs}) = \mathbb{E}_{q(\mathbf{m})}[\log p(\mathbf{m}, \mathbf{d}_{obs})] - \mathbb{E}_{q(\mathbf{m})}[\log q(\mathbf{m})] + \text{KL}[q(\mathbf{m})||p(\mathbf{m}|\mathbf{d}_{obs})]
\label{eq:kl_elbo}
\end{equation}
Since $\log p(\mathbf{d}_{obs})$ is a constant and independent of $q(\mathbf{m})$, minimising KL$[q(\mathbf{m})||p(\mathbf{m}|\mathbf{d}_{obs})]$ is equivalent to maximising the first two terms on the right hand side of equation \ref{eq:kl_elbo}. In addition, since KL$[q||p] \ge 0$, these two terms together act as a lower bound on the logarithmic evidence, and they are usually defined as the \textit{evidence lower bound} (ELBO) of $\log p(\mathbf{d}_{obs})$:
\begin{equation}
\text{ELBO}[q(\mathbf{m})] = \mathbb{E}_{q(\mathbf{m})}[\log p(\mathbf{m}, \mathbf{d}_{obs}) - \log q(\mathbf{m})]
\label{eq:elbo}
\end{equation}
Evaluating ELBO$[q(\mathbf{m})]$ is easier than evaluating KL$[q(\mathbf{m})||p(\mathbf{m}|\mathbf{d}_{obs})]$ since it does not explicitly require the evidence term $p(\mathbf{d}_{obs})$ to be calculated, which is often computationally intractable. The variational problem is therefore often solved by maximising the ELBO$[q(\mathbf{m})]$. The optimisation result is a probability distribution $q^*(\mathbf{m})$ with
\begin{equation}
q^*(\mathbf{m}) = \argmax_{q\in Q}\text{ELBO}[q(\mathbf{m})]
\label{eq:optimal_q}
\end{equation}
which serves as the best approximation to $p(\mathbf{m}|\mathbf{d}_{obs})$ within $\mathcal{Q}(\mathbf{m})$.

In variational inference, there is a trade-off when choosing the variational family: it needs to be sufficiently expressive to provide an accurate approximation to the (potentially complex) posterior distribution, yet simple enough for efficient optimisation. Different choices of the family often result in different variational distributions, and also in different algorithms.

\subsection{Variational Prior Replacement}
Solving the second (quasi) Bayesian problem in equation \ref{eq:bayes_new} requires $p_{new}(\mathbf{m}|\mathbf{d}_{obs})$ to be evaluated. On the right hand side of equation \ref{eq:bayes_new}, while in most cases both old and new prior probability values can be calculated efficiently, the main challenge lies in computing $p_{old}(\mathbf{m}|\mathbf{d}_{obs})$ for a new model $\mathbf{m}$ without invoking equation \ref{eq:bayes_old} (Bayes' rule) which involves forward simulation of data corresponding to $\mathbf{m}$ in order to evaluate the likelihood. Otherwise prior replacement would reduce to solving multiple independent Bayesian inverse problems with different prior distributions. 

We use variational inference to solve the first Bayesian problem described in equation \ref{eq:bayes_old}, providing a known and parametrised probability distribution (called the variational distribution according to the previous section) $q_{old}(\mathbf{m})$ that approximates the old posterior pdf:
\begin{equation}
q_{old}(\mathbf{m}) \approx \ p_{old}(\mathbf{m}|\mathbf{d}_{obs}) =  \dfrac{p(\mathbf{d}_{obs}|\mathbf{m})p_{old}(\mathbf{m})}{p_{old}(\mathbf{d}_{obs})}
\label{eq:bayes_vpr1}
\end{equation}
This distribution is found by solving a variational optimisation problem described in equation \ref{eq:optimal_q}. Once we obtain $q_{old}(\mathbf{m})$, we can use it to replace $p_{old}(\mathbf{m}|\mathbf{d}_{obs})$ in equation \ref{eq:bayes_new}. Since $q_{old}(\mathbf{m})$ is only an approximation (rather than exactly equal) to $p_{old}(\mathbf{m}|\mathbf{d}_{obs})$, we end up with an approximate expression for $p_{new} (\mathbf{m}|\mathbf{d}_{obs})$:
\begin{equation}
p_{new}(\mathbf{m}|\mathbf{d}_{obs}) = k\ p_{old}(\mathbf{m}|\mathbf{d}_{obs})\ \dfrac{p_{new}(\mathbf{m})}{p_{old}(\mathbf{m})} \approx k \ q_{old}(\mathbf{m})\ \dfrac{p_{new}(\mathbf{m})}{p_{old}(\mathbf{m})}
\label{eq:bayes_new_q}
\end{equation}
This means that to estimate $p_{new}(\mathbf{m}|\mathbf{d}_{obs})$ we do not need to evaluate $p_{old}(\mathbf{m}|\mathbf{d}_{obs})$ and so avoid the calculation of likelihood function. Since forward simulation is the most computationally expensive component in an inverse problem, evaluating $q_{old}(\mathbf{m})$ would normally be far cheaper than evaluating $p_{old}(\mathbf{m}|\mathbf{d}_{obs})$. Equation \ref{eq:bayes_new_q} indicates that we can estimate $p_{new}(\mathbf{m}|\mathbf{d}_{obs})$ from $q_{old}(\mathbf{m})$, up to a normalisation constant which can be absorbed into $k$. 

Note that not all variational inference methods can provide a variational distribution ($q_{old}(\mathbf{m})$ in this case) whose probability value can be evaluated easily. For example, Stein variational gradient descent \cite[SVGD --][]{liu2016stein} and its stochastic version \cite[sSVGD --][]{gallego2018stochastic} iteratively update a set of samples (also called particles) such that they become distributed according to an approximation to the posterior distribution. The output is the optimised set of particles, which are used to estimate statistical properties of the posterior distribution. However, evaluating the probability value $q_{old}(\mathbf{m})$ for a particular $\mathbf{m}$ is not at all straightforward with these methods. 

An alternative suite of variational methods approximates the posterior distribution as a known structure with given complexity (for example a Gaussian distribution), and can thus be expressed by a given parametric (often analytic) representation. Variational inference finds the optimal values of hyperparameters that control the parametric expression, thus defining a variational distribution that best approximates the true posterior pdf. Since we obtain a parametric (closed form) expression for the variational distribution, we can easily evaluate its probability value for any model $\mathbf{m}$. We refer to this kind of variational method as \textit{parametric variational inference} \citep{sjolund2023tutorial}. Examples of typical parametric variational inference methods include automatic differentiation variational inference \cite[ADVI --][]{kucukelbir2017automatic}, normalising flows \citep{rezende2015variational}, boosting variational inference \cite[BVI --][]{guo2016boosting, miller2017variational}, and physically structured variational inference \cite[PSVI --][]{zhao2024physically}. 

If $q_{old}(\mathbf{m})$ is constructed using a parametric variational inference method then its probability value can be calculated easily, and equation \ref{eq:bayes_new_q} can in principle be evaluated using any probabilistic inference method since the probability value $p_{new}(\mathbf{m}|\mathbf{d}_{obs})$ can be approximated efficiently. However, even though $p_{new}(\mathbf{m}|\mathbf{d}_{obs})$ can be evaluated efficiently, the curse of dimensionality may nevertheless make the problem expensive, if not impossible, to solve using Monte Carlo sampling methods. We therefore introduce a second variational distribution $q_{new}(\mathbf{m})$ to approximate the new posterior distribution $p_{new}(\mathbf{m}|\mathbf{d}_{obs})$ given the new prior information $p_{new}(\mathbf{m})$. This new variational distribution can be obtained by minimising the KL-divergence between $q_{new}(\mathbf{m})$ and $p_{new}(\mathbf{m}|\mathbf{d}_{obs})$:
\begin{equation}
\small{
\begin{split}
\text{KL}[q_{new}(\mathbf{m})||p_{new}(\mathbf{m}|\mathbf{d}_{obs})] & = \mathbb{E}_{q_{new}(\mathbf{m})}[\log q_{new}(\mathbf{m}) - \log p_{new}(\mathbf{m}|\mathbf{d}_{obs})],\\
& \approx \mathbb{E}_{q_{new}(\mathbf{m})}[\log q_{new}(\mathbf{m}) - \log q_{old}(\mathbf{m}) - \log p_{new}(\mathbf{m}) + \log p_{old}(\mathbf{m})] - \log k
\end{split}}
\label{eq:kl_vpr}
\end{equation}
where the second line is obtained by substituting equation \ref{eq:bayes_new_q} into the first line. Note that $p_{new}(\mathbf{m}|\mathbf{d}_{obs})$ is the exact distribution from equation \ref{eq:bayes_new_q}, which is then approximated by using the same approximation as in equation \ref{eq:bayes_new_q} by introducing the approximate Bayesian solution $q_{old}(\mathbf{m})$. The last term $\log k$ is a constant and can safely be ignored when minimising KL$[q_{new}(\mathbf{m})||p_{new}(\mathbf{m}|\mathbf{d}_{obs})]$. This optimisation problem can be solved in exactly the same way as conventional variational problem, and the result satisfies $q_{new}(\mathbf{m}) \approx p_{new}(\mathbf{m}|\mathbf{d}_{obs})$ obtained by solving
\begin{equation}
q_{new}^*(\mathbf{m}) = \argmin_{q\in Q}\text{KL}[q_{new}(\mathbf{m})||p_{new}(\mathbf{m}|\mathbf{d}_{obs})]
\label{eq:bayes_vpr2}
\end{equation}
Using the framework of variational inference, the two main operations in the original prior replacement problem described in equations \ref{eq:bayes_old} and \ref{eq:bayes_new} are converted into two variational problems to estimate $q_{old}(\mathbf{m})$ and $q_{new}(\mathbf{m})$, containing two approximate steps. We therefore call this new methodology \textit{variational prior replacement} (VPR). 

Note that equation \ref{eq:bayes_vpr1} illustrates that the variational solution to the old Bayesian problem is an approximation, and VPR makes an additional approximation. Even if $q_{old}(\mathbf{m})$ equals $p_{old}(\mathbf{m}|\mathbf{d}_{obs})$ or if we somehow find an exact and analytic solution for $p_{old}(\mathbf{m}|\mathbf{d}_{obs})$, we still need to introduce $q_{new}(\mathbf{m})$ to approximate the true posterior pdf $p_{new}(\mathbf{m}|\mathbf{d}_{obs})$ by minimising the KL divergence KL$[q_{new}(\mathbf{m})||p_{new}(\mathbf{m}|\mathbf{d}_{obs})]$ in equation \ref{eq:kl_vpr} since direct calculation of $p_{new}(\mathbf{m}|\mathbf{d}_{obs})$ using equation \ref{eq:bayes_new_q} requires the normalisation constant $k$ to be evaluated which is intractable in high dimensional inverse problems. While the first operation (equation \ref{eq:bayes_vpr1}) which estimates $q_{old}(\mathbf{m})$ must be performed using a parametric variational method so that its probability value can be evaluated in equations \ref{eq:kl_vpr} and \ref{eq:bayes_vpr2}, the second problem can be solved using any variational method. 

The most expensive step in the VPR algorithm is to solve the first variational problem, because this requires the likelihood term (forward function) to be calculated. Provided that the support of all other prior pdf's are subsets of this support, then this step needs to be performed only once, after which we can replace prior information rapidly whenever it changes. This makes the proposed method attractive in real problems, especially when multiple different prior distributions are possible for a single observed dataset \citep{earp2019probabilistic, bloem2022introducing}.

\subsection{Physically Structured Variational Inference (PSVI)}
\label{section:psvi}
In this paper, we use physically structured variational inference \cite[PSVI --][]{zhao2024physically} to solve the two variational problems to calculate both $q_{old}(\mathbf{m})$ and $q_{new}(\mathbf{m})$. PSVI is an efficient parametric variational inference method that defines a Gaussian variational family with a physics-based correlation structure. When the model parameters to be estimated have physical constrains (for example, seismic velocity should be a positive number and earthquake source location should be below the Earth's surface), a bijective function (an invertible transform) is usually applied to the Gaussian random variables to ensure that the transformed model parameters satisfy their physical constrains. For example, the following \textit{logit} functions
\begin{equation}
\begin{split}
	& m_i = f(\theta_i) = a_i + \dfrac{b_i - a_i}{1+\exp (-\theta_i)}\\
	& \theta_i = f^{-1}(m_i) = \log(m_i - a_i) - \log(b_i - m_i) 
\end{split}
\label{eq:log_transform}
\end{equation}
are often used to convert a Gaussian distributed variable $\theta_i$ defined in an unconstrained space (from minus to plus infinity) into the physical model parameter $m_i$ to be estimated which is bounded by the lower and upper bounds $a_i$ and $b_i$, respectively. The transformed probability distribution can be calculated through the change of variable formula
\begin{equation}
\log p(\mathbf{m}) = \log p(\mathbf{\Theta}) - \log|\det (\partial_{\mathbf{\Theta}} f(\mathbf{\Theta}))|
\label{eq:change_variable}
\end{equation}
where $p(\mathbf{\Theta})$ is the Gaussian variational distribution in the unbounded space. Term $|\det(\cdot)|$ calculates the absolute value of the determinant of the Jacobian matrix $\partial_{\mathbf{\Theta}} f(\mathbf{\Theta})$, which accounts for the volume change corresponding to this transform \citep{kucukelbir2017automatic}.

A Gaussian variational distribution $\mathcal{N}(\boldsymbol\mu, \boldsymbol\Sigma)$ is defined by a mean vector $\boldsymbol\mu$ and a covariance matrix $\boldsymbol\Sigma$. To ensure that $\boldsymbol\Sigma$ always remains positive semi-definite, we re-parametrise it using a Cholesky factorisation $\boldsymbol\Sigma = \mathbf{L}\mathbf{L^T}$, where $\mathbf{L}$ is a lower triangular matrix. A full covariance matrix can be constructed to include correlation information between pairs of model parameters. However, this incurs huge memory requirements and computational costs (for an $n$ dimensional problem, $\mathbf{L}$ requires $n(n+1)/2$ real-valued entries). Alternatively, a mean-field (factorised) Gaussian variational approximation may be used for high dimensional problems, which defines a diagonal covariance matrix, thus ignoring all correlations between model parameters. These two options are respectively referred to as full rank ADVI and mean-field ADVI in \cite{kucukelbir2017automatic}. Unfortunately for the full waveform inversion (FWI) problems considered in this paper, mean-field ADVI normally underestimates uncertainties of the posterior distribution whereas full rank ADVI is intractable due to the dimensionality of models \citep{zhang20233, zhao2024physically}.

PSVI embodies a method with intermediate cost that lies between mean-field ADVI and full rank ADVI, by modelling only the most important (dominant) correlation information in model vector $\mathbf{m}$, guided by physical properties (prior knowledge) of imaging problems. Specifically, in spatial inverse (imaging) problems, model correlations are shown to be strong mainly between pairs of locations that are in spatial proximity to each other, and the magnitude of correlations decreases rapidly as the distance between two locations increases \citep{gebraad2020bayesian, zhang2021bayesian, biswas2022transdimensional}. This suggests that it might be sufficient to model correlations only between parameters that define properties which are spatially close (e.g., for FWI, parameters of cells that lie within a dominant wavelength of one another), and ignore correlations between those that are further apart.

Since off-diagonal elements of the lower triangular matrix $\mathbf{L}$ dominantly represent correlations between parameter pairs, we impose the following sparse structure on $\mathbf{L}$
\begin{equation}
	\mathbf{L} = 
	\begin{bmatrix}
		l_{0,1} &  &  &  &  &  &  \\
		\color{red} l_{1,1} & l_{0,2} &  &  &  &  &  \\
		0 & \color{red} l_{1,2} & l_{0,3} &  &  &  &  \\
		... & 0 & \color{red} l_{1,3} & ... &  &  &  \\
		\color{red} l_{i,1} & ... & 0 & \color{red} ... & l_{0,n-2} &  &  \\
		0 & \color{red} ... & ... & ... & \color{red} l_{1,n-2} & l_{0,n-1} &  \\
		... & 0 & \color{red} l_{i,n-i} & ... & 0 & \color{red} l_{1,n-1} & l_{0, n} 
	\end{bmatrix}
	\label{eq:structured_l}
\end{equation}
For each element, the first subscript $i$ indicates a block of off-diagonal elements that are $i$ rows below the main diagonal (i.e., at an offset of $i$ from the main diagonal), and the second subscript $j$ indicates that $l_{i,j}$ is the $j$th element of that off-diagonal block. In equation \ref{eq:structured_l}, sparsely distributed off-diagonal elements in red are used to capture main correlations between parameter pairs that are assumed to be important (in this case, are spatially close), and their values are optimised during variational inference. All other off-diagonal elements of $\mathbf{L}$ are set to be zero by assuming independence of the corresponding model parameter pairs. Note that we only impose a sparse structure on $\mathbf{L}$ rather than setting constraints on the values of the non-zero off-diagonal elements in red: those values are updated freely during the variational optimisation \citep{zhao2024physically}.

In PSVI, we can impose any desired correlation structure on $\mathbf{L}$ by setting only the corresponding off-diagonal blocks as unknown hyperparameters and optimising them. The total number of parameters to define $\mathbf{L}$ can thus be greatly reduced compared to that in full rank ADVI. The covariance matrix $\boldsymbol\Sigma$ obtained in this way then also represents a sparse correlation structure with specific non-zero off-diagonal blocks (similar to the red elements in equation \ref{eq:structured_l} but located below and above the main diagonal elements). Since a \textit{priori} we expect that most of the important correlations are included in PSVI, the obtained variational distribution would capture parameter correlations of interest. Thus, inference results are significantly improved compared to those from mean-field ADVI \citep{zhao2024physically}.

Variational parameters $\boldsymbol\mu$ and $\mathbf{L}$ are updated by maximising the EBLO in equation \ref{eq:elbo} or equivalently minimising the KL divergence in equations \ref{eq:kl} and \ref{eq:kl_vpr} using gradient based optimisation methods, and their gradients are calculated using automatic differentiation libraries \citep{abadi2016tensorflow, paszke2019pytorch}. The expectation terms are estimated by Monte Carlo integration with a relatively small number of samples drawn from the variational distribution, because the optimisation is performed over many iterations so that statistically the parameters will converge towards the correct solution \citep{kucukelbir2017automatic}.

\section{Application to Full Waveform Inversion}
Seismic full waveform inversion (FWI) estimates subsurface physical properties, such as seismic velocities and density, using seismic waveform data \citep{tarantola1984inversion, fichtner2009full, virieux2009overview}. FWI is a highly nonlinear and non-unique inverse problem, and thus deterministic methods often fail to find a truly representative Earth model that generates the observed data, and to estimate reliable uncertainties in the inversion results. In recent years, researchers have started to use various Bayesian inference methods to solve probabilistic FWI problems, including Monte Carlo sampling methods \citep{ray2016frequency, ray2018low, visser2019bayesian, gebraad2020bayesian, guo2020bayesian, kotsi2020uncertainty, zhao2021gradient, khoshkholgh2022full, biswas2022transdimensional, fu2022time, de2023acoustic, de2023bayesian,  berti2023computationally} and variational inference \citep{zhang2021bayesianfwi, bates2022probabilistic, wang2023re, lomas20233d, izzatullah2023physics, zhao2024physically, yin2024wise}. Bayesian FWI often requires huge computational resources since (1) the dimensionality (number of unknown parameters) of an FWI problem is usually high \citep{curtis2001prior}, and (2) the forward and adjoint simulations are expensive \citep{wang2019cu, zhao2020domain}. Therefore, reducing computational overhead (while still obtaining reasonably accurate inversion results) is a top priority in Bayesian FWI, especially if we have multiple prior distributions to consider.

\begin{figure}
	\centering\includegraphics[width=\textwidth]{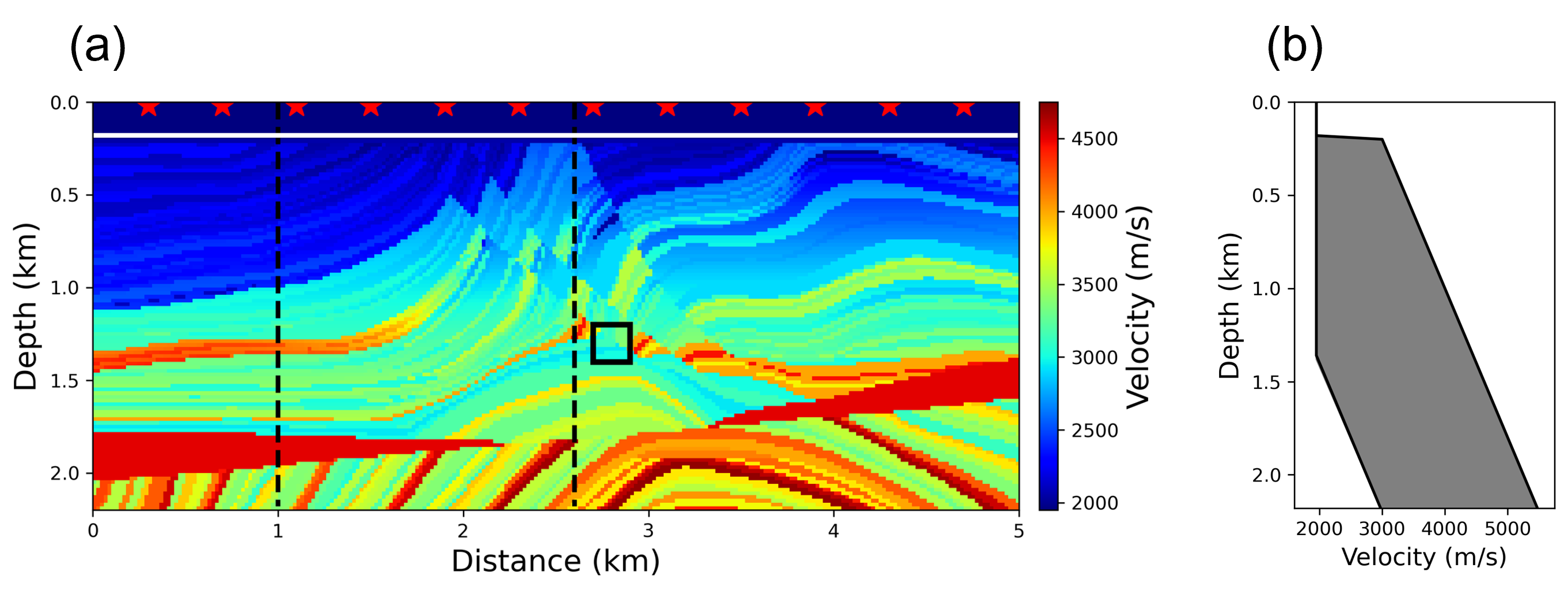}
	\caption{(a) Truncated and down-sampled P wave velocity of the Marmousi model used in this paper. Source locations are indicated by red stars and the receiver line is marked by a white line. Dashed black lines display the locations of two vertical profiles used to compare the posterior marginal probability distributions in Figures \ref{fig:test_vpr_marginal} and \ref{fig:fwi_marginal} in the main text. (b) Upper and lower bounds for the uniform prior distribution at different depths.}
	\label{fig:fwi_110_250_vel_data_prior}
\end{figure}

We apply variational prior replacement (VPR) to a 2D Bayesian acoustic FWI problem to explore its effectiveness. Figure \ref{fig:fwi_110_250_vel_data_prior}a displays the true velocity model used in the following tests, which is obtained by truncating and downsampling the original Marmousi model \citep{martin2006marmousi2} to a grid size of 110 $\times$ 250 cells, with each cell measuring 20 m in both horizontal and vertical directions. We place 12 sources (red stars in Figure \ref{fig:fwi_110_250_vel_data_prior}a) on the surface with a spacing of 400 m, and 250 receivers (white line in Figure \ref{fig:fwi_110_250_vel_data_prior}a) on the seabed (200 m depth) with a horizontal interval of 20 m. The waveform data are 4 s long with a sample interval of 2 ms, which are generated by solving a 2D acoustic wave equation using a time-domain finite difference method. We further add Gaussian random noise with zero mean and a standard deviation value of 0.1 ($\sim 1\%$ of the average value of the maximum amplitude of each seismic trace) to the obtained waveform data, which is treated as the observed dataset in this example. The source function is a Ricker wavelet with a dominant frequency of 10 Hz.

We define a Gaussian likelihood function to represent data uncertainties 
\begin{equation}
p(\mathbf{d}_{obs}|\mathbf{m}) \propto \exp \left[-\dfrac{(\mathbf{d}_{syn}-\mathbf{d}_{obs})^T \Sigma_\mathbf{d}^{-1} (\mathbf{d}_{syn}-\mathbf{d}_{obs})}{2}\right]
\label{eq:likelihood}
\end{equation}
In this example, the covariance matrix of data noise $\Sigma_{\mathbf{d}}$ is assumed to be a diagonal matrix (uncorrelated data noise), and all diagonal elements are set to be 0.1 to represent the level of noise added to the synthetic waveform data. That is to say, we assume the data covariance matrix $\Sigma_{\mathbf{d}}$ is known. The same finite difference forward modelling method is used to calculate synthetic data $\mathbf{d}_{syn}$, and data-model gradients are computed using the adjoint-state method \citep{plessix2006review}. During forward and adjoint simulations, we fix velocity values in the water layer at their true values. Prior distributions used in this study are discussed below.

\subsection{Prior information}
We consider three different types of prior information in the FWI problem. We first define a uniform prior distribution $p_1(\mathbf{m})$ for the velocity values at each grid cell with lower and upper bounds at each depth displayed in Figure \ref{fig:fwi_110_250_vel_data_prior}b, similar to that used in \cite{zhang2021bayesianfwi}. This is a non-informative (weak) and thus broad prior distribution with no correlations between neighbouring cells. It has the advantage that any type of velocity contrast between neighbouring cells would be consistent with this prior pdf (all model samples have the same prior probability density) as long as they lie within the prior bounds, and hence can in principle be discriminated only by comparing their consistency with the observed waveform data.

The second prior distribution is a spatially smoothed version of the uniform distribution, obtained by applying a second-order finite difference (smoothing) operator $\mathbf{S}$:
\begin{equation}
\mathbf{S} = 
	\begin{bmatrix}
	 &  & & ... &  &  &  \\
	...& 1 & -2 & 1 & 0 & 0 & ...\\
	...& 0 & 1 & -2 & 1 & 0 & ...  \\
	...& 0 & 0 & 1 & -2 & 1 & ... \\
	&  &  & ... &  &  &  
	\end{bmatrix}
\label{eq:smooth_operator}
\end{equation}
to model parameter $\mathbf{m}$. Define a Gaussian distribution for $\mathbf{Sm}$
\begin{equation}
p(\mathbf{Sm}) = k_1 \exp \left[-\frac{1}{2}(\mathbf{Sm})^T \Sigma_{\mathbf{Sm}}^{-1} (\mathbf{Sm})\right]
\label{eq:sm}
\end{equation}
where $k_1$ is a normalisation constant, and $\Sigma_{\mathbf{Sm}}$ is a diagonal matrix with its diagonal elements controlling the strength of the spatial smoothness (larger values correspond to weaker spatial smoothness). In this paper, the diagonal elements of $\Sigma_{\mathbf{Sm}}$ are set to 500. This can be interpreted as applying a Tikhonov (regularisation) matrix $\mathbf{S}$ to $\mathbf{m}$ \citep{golub1999tikhonov, aghamiry2018hybrid}. Then the smoothed prior distribution $p_2(\mathbf{m})$ can be written as
\begin{equation}
p_2(\mathbf{m}) = \dfrac{p(\mathbf{Sm})p_1(\mathbf{m})}{k_2}
\label{eq:smooth_prior}
\end{equation}
where $p_1(\mathbf{m})$ is the uniform distribution defined above, and $k_2$ is another normalisation constant which can be absorbed into the evidence term in Bayes' rule so we do not need to calculate its value.

This prior distribution $p_2(\mathbf{m})$ embodies strong prior information in which model samples with smaller velocity contrasts between spatially neighbouring cells have higher probability values. Therefore, velocities in neighbouring cells should be positively correlated. This information may or may not be advantageous depending on the true (geological) prior information about the form of the velocity structure being estimated. Compared to the uniform prior distribution, large velocity contrasts between neighbouring cells are almost excluded by this prior information as they have relatively low prior probability values. This effectively reduces the hypervolume of parameter space spanned by significantly non-zero values of the posterior pdf. In other words, it provides more information than $p_1(\mathbf{m})$. More detailed comparisons of these two prior distributions and their effects on the posterior pdfs can be found in \cite{earp2019probabilistic}.

The third prior distribution $p_3(\mathbf{m})$ is a Gaussian distribution with real geology-informed inter-parameter correlation information. The mean and standard deviation vectors of the Gaussian prior distribution are set to be those of the uniform prior distribution $p_1(\mathbf{m})$ for consistency. Considering the dimensionality of this FWI problem (100 $\times$ 250), it is difficult to build a full prior covariance matrix to describe detailed geological prior information. We therefore build a prior covariance matrix that incorporates only a local correlation structure estimated from real geology.

\begin{figure}
	\centering\includegraphics[width=0.5\textwidth]{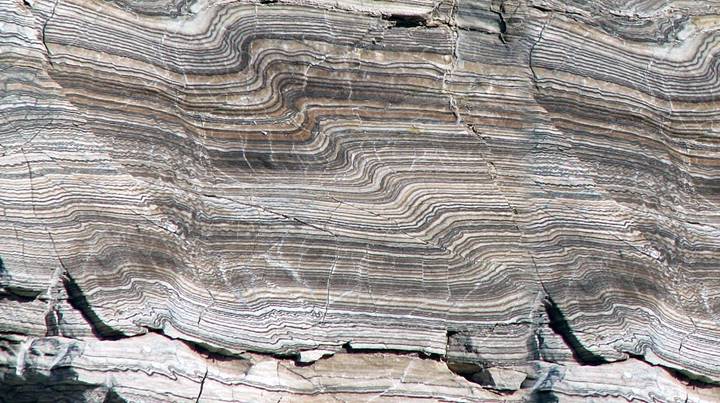}
	\caption{A picture of real geological structures with a scale of metres \citep{james2017folded} used to calculate a local correlation matrix and define the geological prior distribution $p_3(\mathbf{m})$.}
	\label{fig:geological_images}
\end{figure}

\begin{figure}
	\centering\includegraphics[width=\textwidth]{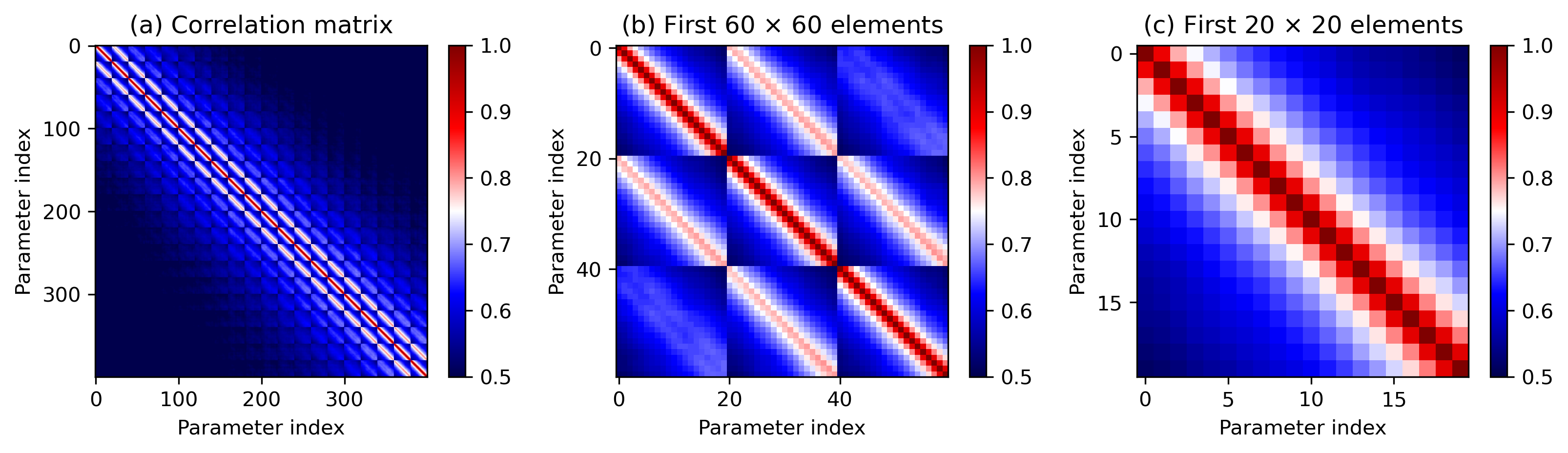}
	\caption{(a) Prior correlation matrix calculated from a set of 20 $\times$ 20 sized training images sampled from pictures of real geology such as Figure \ref{fig:geological_images}. (b) and (c) show magnifications of the first 60 $\times$ 60 and 20 $\times$ 20 elements in (a), respectively. Due to parametrisation of the training images, distinct off-diagonal blocks below and above the main diagonal block in (a) and (b) represent vertical correlations, and off-diagonals directly below and above the main diagonal elements in (c) denote horizontal correlations.}
	\label{fig:prior_cor_geological}
\end{figure}

To achieve this, we first select a set of realistic geological images. Figure \ref{fig:geological_images} displays one such image \citep{james2017folded}, and while the images are all at much smaller scale than the Marmousi model was designed to represent, we assume scale invariance of geological correlations (only for the purposes of this test). From each of the pictures, we randomly sample 1000 subimages using a window with 20 $\times$ 20 pixels, which together represent a local correlation structure between parameters. Figure \ref{fig:prior_cor_geological}a shows the calculated full correlation matrix with a size of 400 $\times$ 400, each element denoting correlation information of one parameter pair within the 20 $\times$ 20 window. To analyse a more detailed structure of this correlation matrix, Figures \ref{fig:prior_cor_geological}b and \ref{fig:prior_cor_geological}c present its first 60 $\times$ 60 elements and 20 $\times$ 20 elements, respectively. Note that we reshape the 20 $\times$ 20 (2D) images into 1D vectors in a row-major order (i.e., for each training image the first 20 elements of the 1D vector comprise the first row of the 2D image, the second 20 elements comprise the second row, and so on). Therefore, off-diagonal blocks observed in Figures \ref{fig:prior_cor_geological}a and \ref{fig:prior_cor_geological}b represent correlation information in the vertical direction; only one such obvious block is visible, meaning that vertical correlations exist predominantly between vertically adjacent cells and decay rapidly with greater inter-cell distance. On the other hand, off-diagonals directly below and above the main diagonal elements (displayed in Figure \ref{fig:prior_cor_geological}c) denote horizontal correlations: three or four strong off-diagonal elements have correlation values larger than 0.7, implying that strong horizontal correlations exist within approximately 4 neighbouring cells. Such horizontally smoother and vertically rougher correlation features are also clear in Figure \ref{fig:geological_images}.

We use this correlation matrix to construct a full correlation matrix $\mathbf{R}$ that describes correlations in model vector $\mathbf{m}$ (with a dimensionality of 100 $\times$ 250 in this example) by considering correlations between pairs of parameters that are located only inside a 20 $\times$ 20 window. We set all other elements to zero for reasons discussed in Section \ref{section:psvi} and in \cite{zhao2024physically}. The covariance matrix of this Gaussian prior distribution $\boldsymbol{\Sigma}_{p_3}$ can then be calculated by
\begin{equation}
	\boldsymbol{\Sigma}_{p_3} = \mathbf{D}_{std}\ \mathbf{R}\ \mathbf{D}_{std}
\label{eq:cov_gaussian_prior}
\end{equation}
where $\mathbf{D}_{std}$ is a diagonal matrix with diagonal elements being the standard deviations of $p_3(\mathbf{m})$, and $\mathbf{R}$ is the correlation matrix obtained above. Finally, the Gaussian prior distribution $p_3(\mathbf{m})$ can be defined as
\begin{equation}
p_3(\mathbf{m}) = k_3 \exp \left[-\dfrac{1}{2}(\mathbf{m} - \boldsymbol{\mu}_{p_3})^T \boldsymbol{\Sigma}_{p_3}^{-1} (\mathbf{m} - \boldsymbol{\mu}_{p_3})\right]
\label{eq:gaussian_prior}
\end{equation}
where $\boldsymbol{\mu}_{p_3}$ is the mean vector of this Gaussian distribution. Similarly to equation \ref{eq:smooth_prior}, $k_3$ is a normalisation constant whose value is not required in VPR. Below, $p_3(\mathbf{m})$ is referred to as the geological prior distribution since it captures spatial correlation information from real geological structures.

\begin{figure}
	\centering\includegraphics[width=0.6\textwidth]{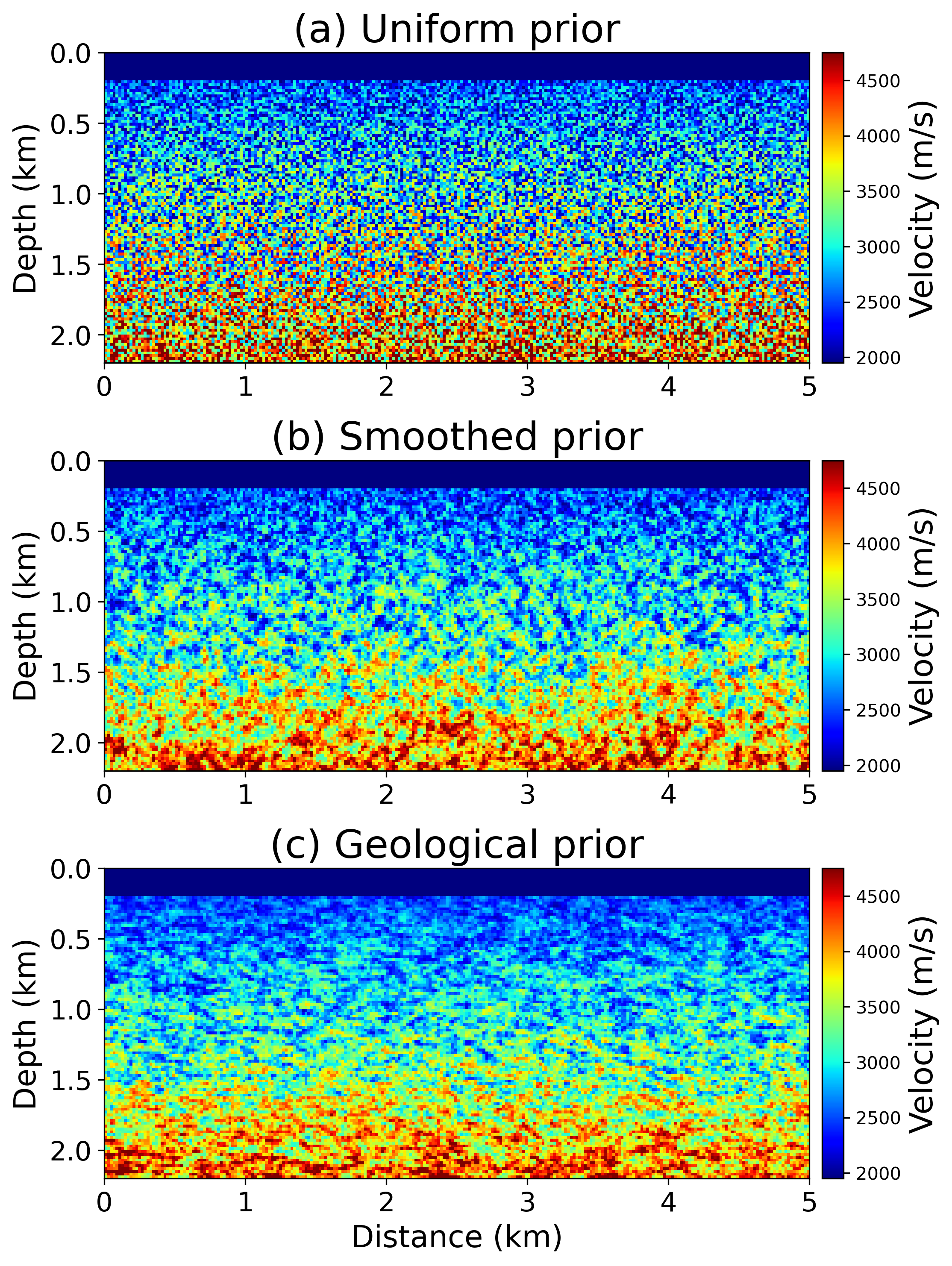}
	\caption{(a) -- (c) One random prior sample drawn from the (a) uniform, (b) smoothed and (c) geological prior distributions defined in the main text, respectively.}
	\label{fig:prior_samples}
\end{figure}

Figure \ref{fig:prior_samples} displays one random sample drawn from each of the three prior distributions. Since no spatial correlation is considered in the uniform prior distribution, we observe large velocity contrasts between neighbouring cells in Figure \ref{fig:prior_samples}a. The smoothed and geological prior distributions impose spatially correlated information, thus the prior samples presented in Figures \ref{fig:prior_samples}b and Figures \ref{fig:prior_samples}c are spatially smoother. In addition, local velocity structures in Figure \ref{fig:prior_samples}c show rectangular patterns with larger sizes in the horizontal direction and smaller sizes in the vertical direction due to horizontal smoothness and vertical roughness as represented in the geological prior pdf and illustrated in Figures \ref{fig:geological_images} and \ref{fig:prior_cor_geological}. This pattern is not observed in Figure \ref{fig:prior_samples}b since in $p_2(\mathbf{m})$ we impose the same magnitude of smoothness in both horizontal and vertical directions (this was not a requirement, but in equation \ref{eq:sm} we used equal horizontal and vertical smoothing to contrast with the geological prior pdf). Figure \ref{fig:prior_samples} thus proves that the three prior distributions encapsulate significantly different prior information that we may wish to inject into FWI inversion results.

\subsection{Verifying variational prior replacement}
\label{section:verify_vpr}

In the first test, we verify that VPR produces correct results by comparing them to those obtained using a conventional approach where an independent Bayesian inversion is performed for each prior distribution (referred to herein as \textit{prior specific inversion}). For this test, we consider the uniform prior distribution $p_1(\mathbf{m})$ and the smoothed prior $p_2(\mathbf{m})$. For the prior specific case, PSVI is used to solve these two FWI problems with their respective priors. We update variational parameters (mean vector $\boldsymbol\mu$ and lower triangular matrix $\mathbf{L}$ mentioned in Section \ref{section:psvi}) for 5,000 iterations. During each iteration, 2 random samples are employed to approximate the ELBO$[q(\mathbf{m})]$ (equation \ref{eq:elbo}) using Monte Carlo integration; such a low number of samples has been shown to be reasonable in a stochastic sense in previous studies, because of the large number of iterations \citep{kucukelbir2017automatic}. For VPR, the uniform distribution $p_1(\mathbf{m})$ is treated as the old prior distribution, which is then removed from the old posterior distribution (the posterior pdf calculated using $p_1(\mathbf{m})$ by prior specific inversion) and replaced by the smoothed (new) prior distribution $p_2(\mathbf{m})$. This is achieved by solving the variational problem described in equation \ref{eq:bayes_vpr2}, through minimisation of the KL divergence expressed in equation \ref{eq:kl_vpr}. Similarly to the prior specific case, PSVI is used to solve this problem, where variational parameters are updated for 5,000 iterations with 10 samples per iteration used to estimate the expectation term in equation \ref{eq:kl_vpr}. Note that it might be impossible to replace a smoothed (old) prior by a Uniform (new) prior distribution using VPR since in this case the support of the old prior may only be a subset of that of the new prior distribution (or it might be effectively so due to sampling and numerical approximations). This might make $p_{new}(\mathbf{m})/p_{old}(\mathbf{m})$ numerically unstable because for some parameters $\mathbf{m}$ the value $p_{old}(\mathbf{m})$ could be small, poorly determined or even zero.

\begin{figure}
	\centering\includegraphics[width=\textwidth]{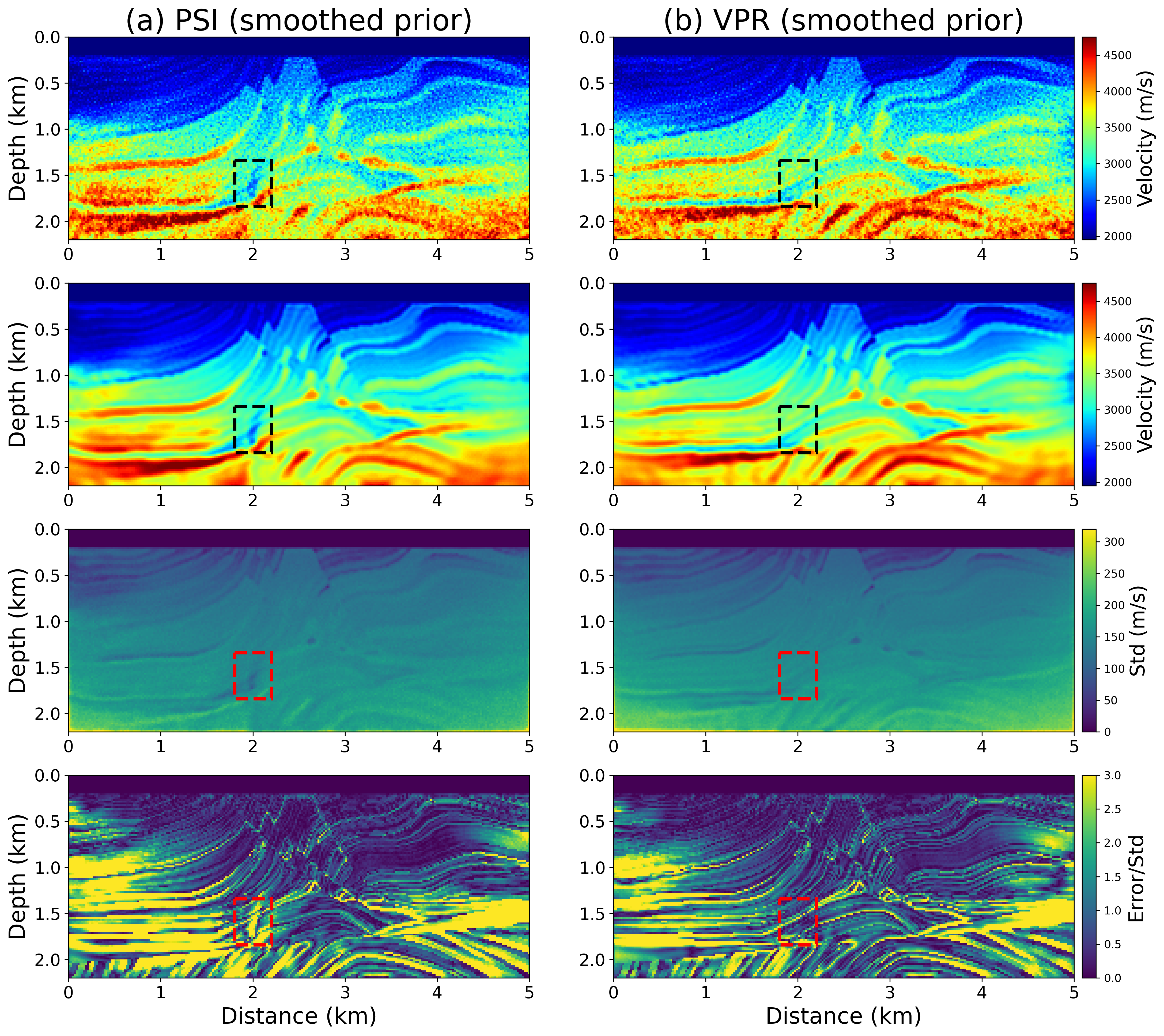}
	\caption{(a) Prior specific inversion (PSI) and (b) variational prior replacement (VPR) results obtained using the smoothed prior distribution $p_2(\mathbf{m})$. The latter is obtained by removing the uniform prior information $p_1(\mathbf{m})$ from the old posterior pdf (displayed in Figure \ref{fig:fwi_mean_std}a), and imposing the smoothed prior $p_2(\mathbf{m})$ using VPR. From top to bottom row, they are: a random posterior sample, mean velocity map, standard deviation and relative error of the obtained posterior distribution, respectively. The relative error is the absolute error between the mean and true models divided by the corresponding standard deviation at each point. Red and black dashed boxes highlight differences between (a) and (b).}
	\label{fig:test_vpr_mean}
\end{figure}

Figures \ref{fig:test_vpr_mean}a and \ref{fig:test_vpr_mean}b display prior specific inversion (PSI) and VPR results obtained using the smoothed prior distribution. In each column, a random posterior sample, the mean velocity map, standard deviation and the relative error of the posterior distribution are presented from top to bottom row. The relative error defined to be the difference between the mean and true velocity models (Figure \ref{fig:fwi_110_250_vel_data_prior}a) divided by the standard deviation at each point, reflecting the relative deviation between the true and inverted mean models. The most important feature is that the first-order posterior statistics displayed in Figures \ref{fig:test_vpr_mean}a and \ref{fig:test_vpr_mean}b are almost identical; Figure \ref{fig:test_vpr_mean}b was produced from the results obtained using the uniform prior (displayed in Figure \ref{fig:fwi_mean_std}a). This supports the statement that VPR is able to replace the old prior information and inject (update) the new prior information into the inversion results, without solving a Bayesian inverse problem again from scratch. 

We do observe some discrepancies between these two sets of results. For example, a vertically oriented low velocity structure is present in Figure \ref{fig:test_vpr_mean}a inside dashed black and red boxes, which is not present in Figure \ref{fig:test_vpr_mean}b. Although this feature is not observed in the true velocity model in Figure \ref{fig:fwi_110_250_vel_data_prior}a, there is no strong evidence to discriminate which result is better. 

\begin{figure}
	\centering\includegraphics[width=0.7\textwidth]{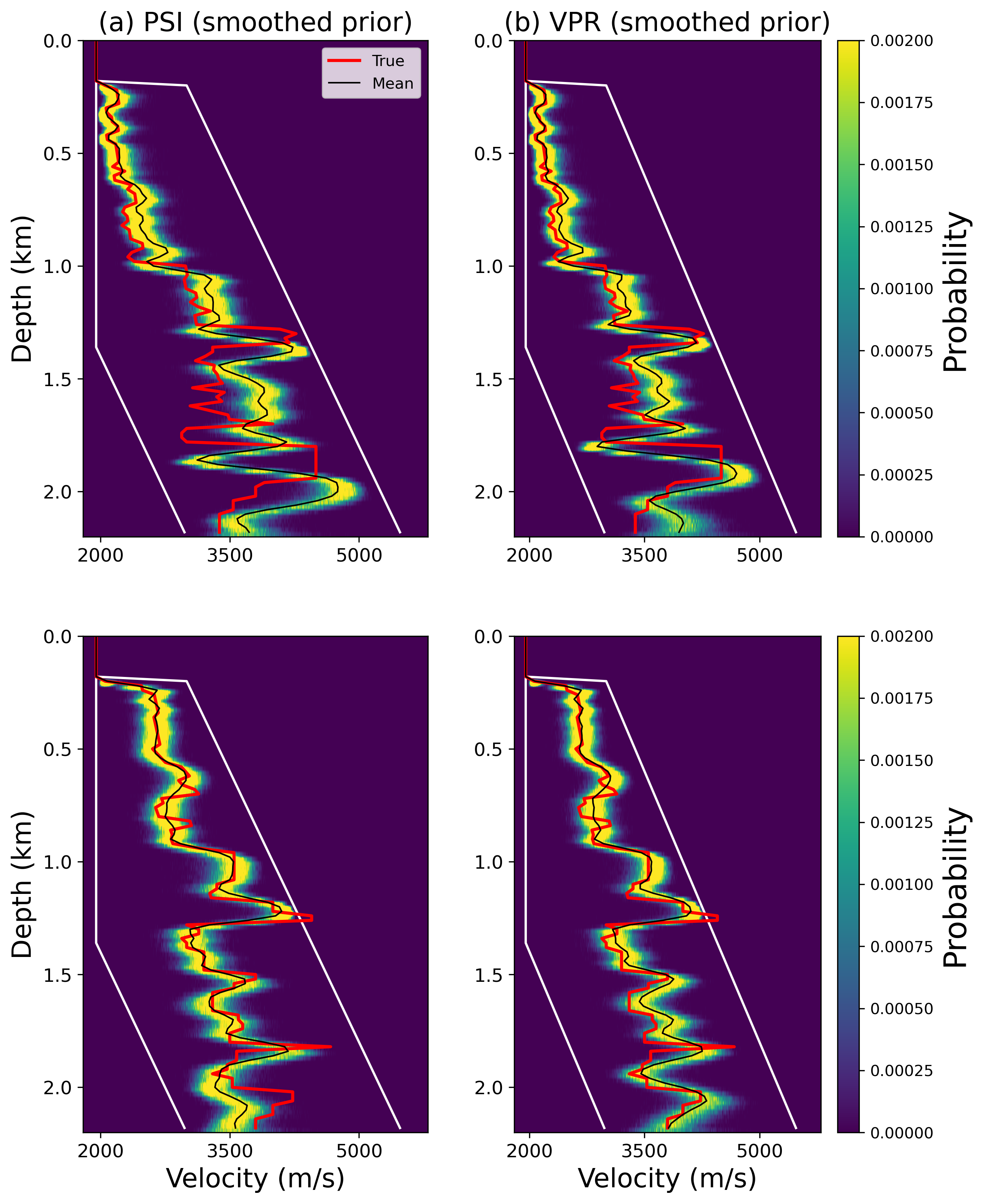}
	\caption{Posterior marginal distributions coloured from dark blue (zero probability) to yellow (maximum value of marginal pdf's in each plot), along two vertical velocity profiles at horizontal locations of 1 km (top row) and 2.6 km (bottom row) whose locations are marked by black dashed lines in Figure \ref{fig:fwi_110_250_vel_data_prior}a. (a) and (b) Posterior marginal pdfs from prior specific inversion (PSI) and variational prior replacement (VPR), obtained using the smoothed prior distribution.}
	\label{fig:test_vpr_marginal}
\end{figure}

For other local regions such as those below and to the right of the boxes in Figure \ref{fig:test_vpr_mean}, the mean velocity model from VPR (Figure \ref{fig:test_vpr_mean}b) appears to be closer to the old posterior distribution using the uniform prior pdf (displayed in Figure \ref{fig:fwi_mean_std}a) than it does to the PSI result using the smoothed prior (Figure \ref{fig:test_vpr_mean}a). This might be because this implementation of PSI (which is an optimisation problem) converged to a local minimum, or that it has not fully converged (full convergence might require a larger number of forward evaluations which is very expensive). Alternatively, it is also possible that the VPR procedure (also an optimisation process) has not fully converged and thus might indicate an incomplete prior replacement in this test. On the other hand, for some other statistics such as the characteristic of spatial variations in each posterior sample, standard deviations, or posterior marginal pdfs displayed in Figure \ref{fig:test_vpr_marginal}, VPR results are clearly closer to the PSI results using the smoothed prior than to those using the uniform prior, supporting the statement that VPR produces reasonable statistical accuracy. We also note that it is reasonable that there remain some discrepancies, caused by the fact that in VPR we introduce a variational distribution $q_{new}(\mathbf{m})$ to approximate the new posterior distribution $p_{new}(\mathbf{m}|\mathbf{d}_{obs})$ as expressed in equation \ref{eq:bayes_vpr2}, rather than calculating the actual $p_{new}(\mathbf{m}|\mathbf{d}_{obs})$ as in \cite{walker2014varying}.

\begin{figure}
	\centering\includegraphics[width=0.75\textwidth]{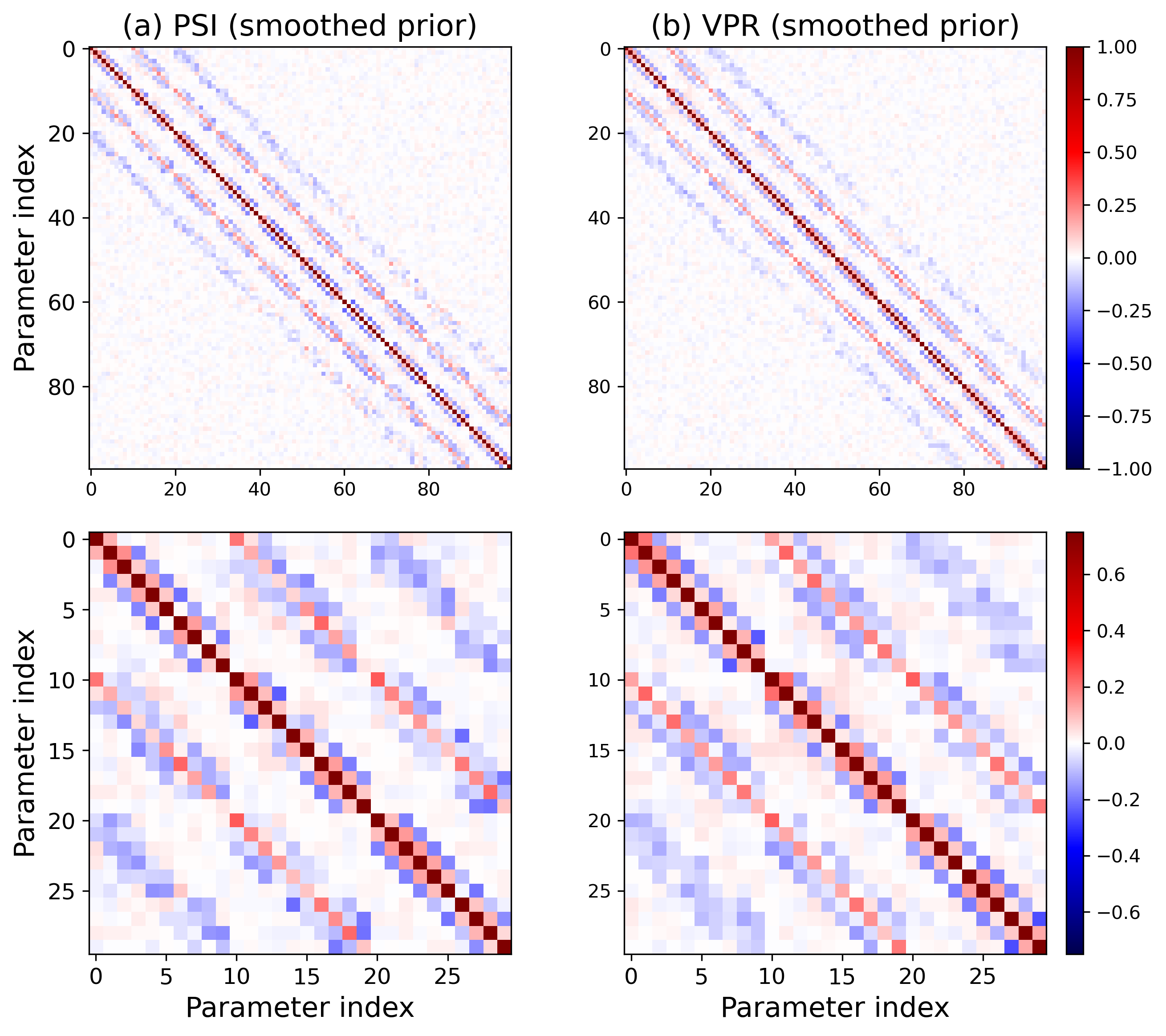}
	\caption{Posterior correlation matrices from (a) PSI and (b) VPR results for velocity values in a 10 $\times$ 10 window inside the black box in Figure \ref{fig:fwi_110_250_vel_data_prior}a. Top row shows the full 100 $\times$ 100 sized posterior correlation matrices and bottom row shows the first 30 $\times$ 30 elements for better comparison.}
	\label{fig:test_vpr_correlation}
\end{figure}

Figure \ref{fig:test_vpr_marginal} compares the posterior marginal pdfs of the above two results along two vertical velocity profiles at horizontal locations of 1 km (top row) and 2.6 km (bottom row). Their locations are displayed by dashed black lines in Figure \ref{fig:fwi_110_250_vel_data_prior}a. Red and black lines represent the true and mean velocity values, respectively. Despite some small discrepancies here and there in Figures \ref{fig:test_vpr_marginal}a and \ref{fig:test_vpr_marginal}b, they provide similar posterior marginal pdfs. Interestingly, as displayed in the top row, the two methods (PSI and VPR) find very similar yet \textit{incorrect} posterior solutions given the same data and prior information (especially below 1.3 km depth where true velocity values are excluded by the high probability region of the posterior pdfs). This is because that the PSVI algorithm may have converged around an incorrect solution in this region caused by cycle skipping, which often occurs in FWI problems. We also compare correlation information of the two posterior pdfs in Figure \ref{fig:test_vpr_correlation}, which displays the posterior correlation matrices for velocity values in a 2D window with a size of 10 $\times$ 10 cells inside the black box in Figure \ref{fig:fwi_110_250_vel_data_prior}a. The top row shows the full correlation matrices (with a size of 100 $\times$ 100), and the bottom row shows the first 30 $\times$ 30 elements. Highly consistent posterior correlation values are obtained, which again proves the effectiveness of VPR.

\begin{figure}
	\centering\includegraphics[width=0.75\textwidth]{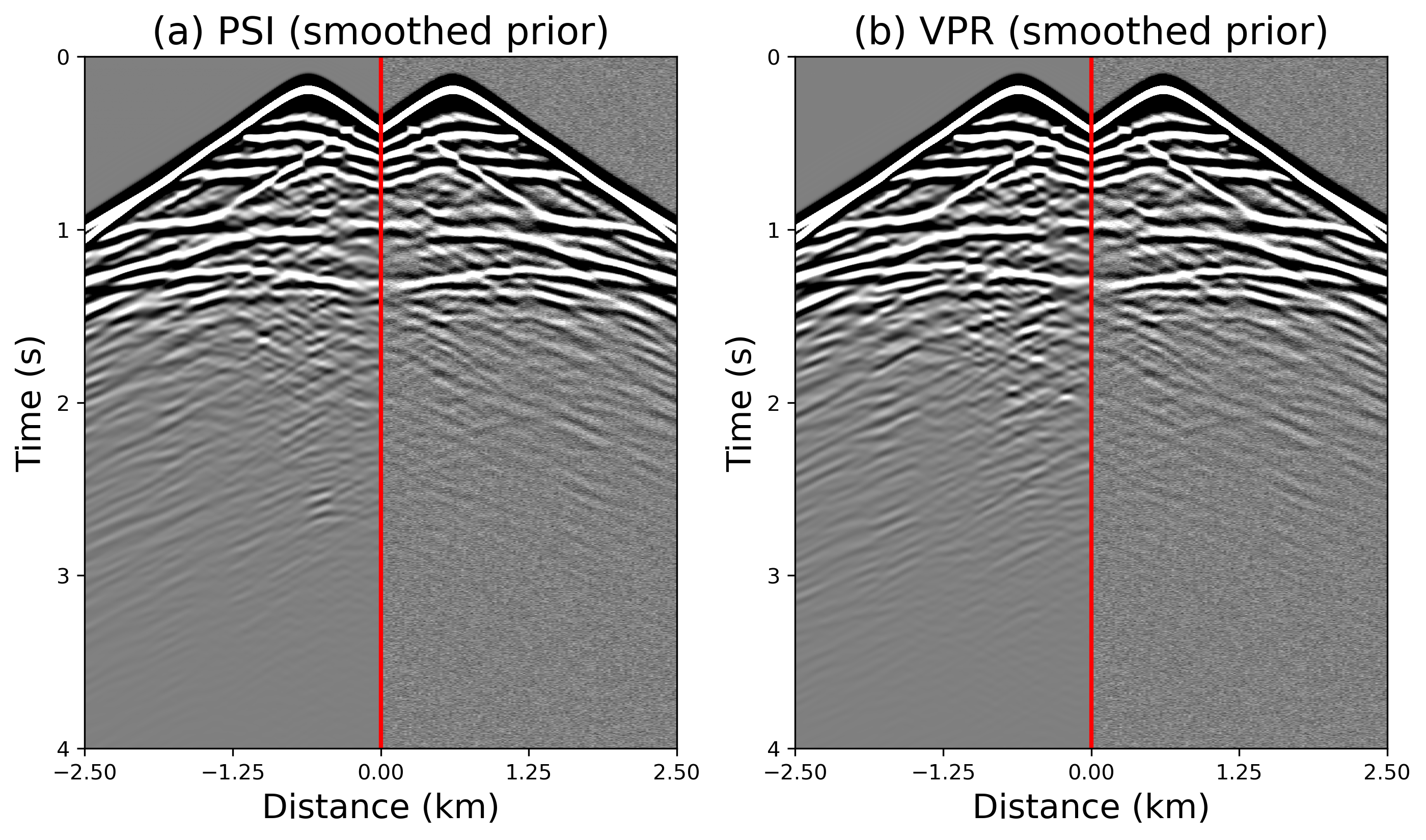}
	\caption{The ``butterfly plot'' of data comparison for one common shot gather. (a) The data predicted by a random posterior sample from PSI (left hand side of red line) and the observed waveform data (right hand side of red line). (b) The data predicted by a posterior sample from VPR (left) and the observed waveform data (right). In both figures, the simulated data are highly consistent with the observed (noisy) data.}
	\label{fig:fwi_110_250_datafit_vpr_psi}
\end{figure}

To further test the performance of VPR, in Figures \ref{fig:fwi_110_250_datafit_vpr_psi}a and \ref{fig:fwi_110_250_datafit_vpr_psi}b we compare one observed shot gather with data simulated by one randomly chosen posterior sample obtained from PSI and VPR results, respectively. In both figures, the simulated data are highly consistent with the observed (noisy) data, which demonstrates that VPR can provide models that produce accurate waveform data that fit the observed data to within data uncertainties.
	
In Appendix \ref{ap:verify_vpr}, we present a second example to test the accuracy of VPR using a Gaussian prior distribution for $p_1(\mathbf{m})$, which again shows that VPR and PSI provide highly consistent inversion results and posterior statistics. In conclusion, since variational prior replacement and prior specific inversion provide almost identical posterior random samples, first-order statistics (posterior mean, standard deviation and marginal pdfs) and second-order statistics (correlation matrices), we assert that the proposed method is effective and accurate for varying prior information in Bayesian inference without performing repeated independent inversions.

\subsection{FWI using different priors}
\label{section:3priors}

In this section we analyse the effect of the three different priors defined previously and compare the corresponding inversion results. We employ PSVI to perform a single variational Bayesian FWI using the uniform distribution $p_1(\mathbf{m})$. $p_1(\mathbf{m})$ is then replaced by both the smoothed prior $p_2(\mathbf{m})$ and the geological prior $p_3(\mathbf{m})$ using VPR. Note that in VPR the old prior distribution should be broader than the new prior to avoid numerical instability issues such as occur when attempting to divide by zero \citep{walker2014varying}. However, the Gaussian geological prior distribution is defined in the space of real numbers, which spans a broader parameter space than the uniform distribution. Therefore, we truncate $p_3(\mathbf{m})$ within the lower and upper bounds of the uniform prior distribution, and renormalise $p_3(\mathbf{m})$ by another normalisation constant (which does not need to be evaluated, similar to those in equations \ref{eq:smooth_prior} and \ref{eq:gaussian_prior}).

\begin{figure}
	\centering\includegraphics[width=\textwidth]{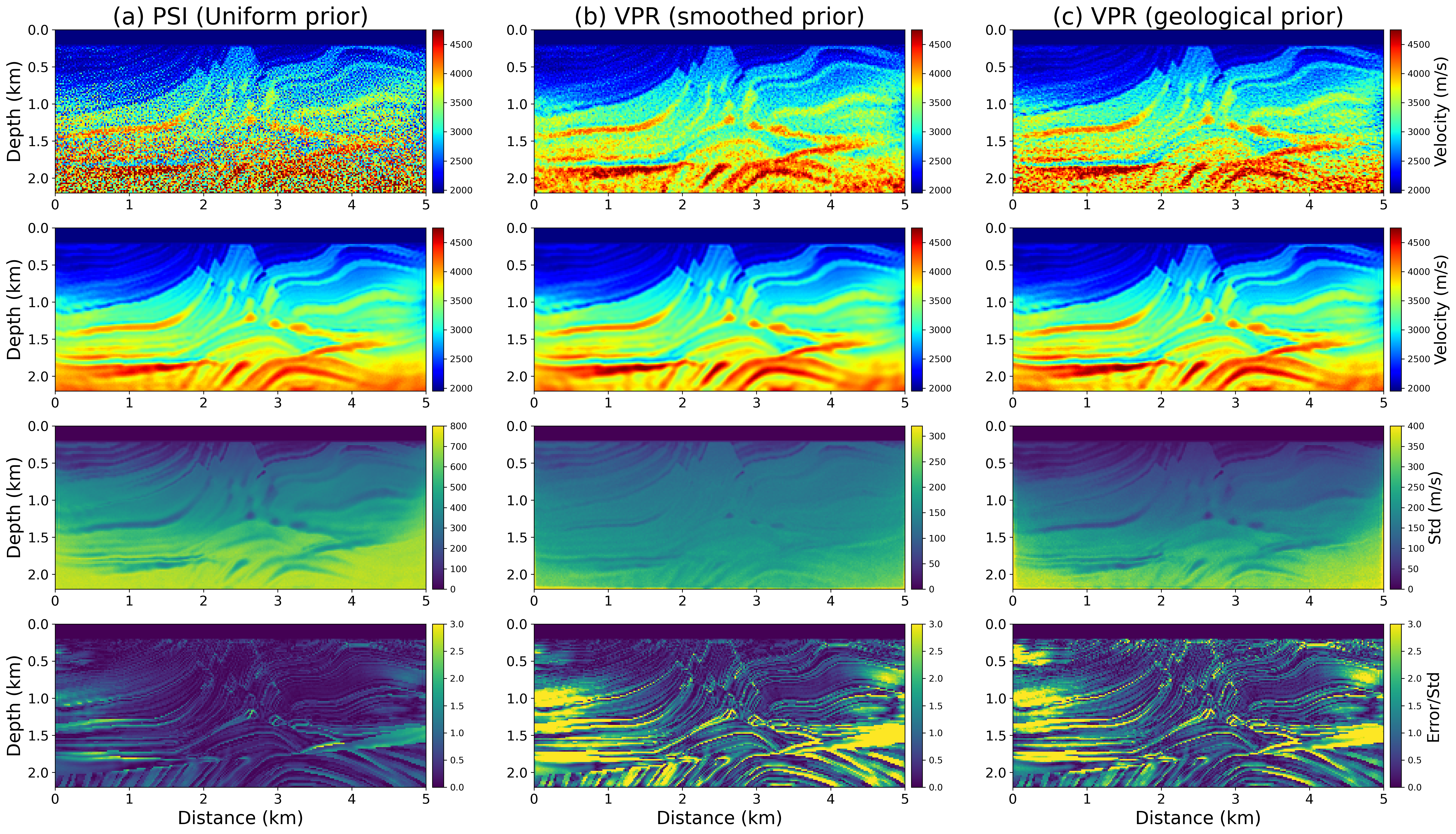}
	\caption{(a) Prior specific inversion (PSI) results obtained using the uniform prior distribution $p_1(\mathbf{m})$. (b) and (c) Variational prior replacement (VPR) results obtained by replacing the uniform prior distribution $p_1(\mathbf{m})$ by the smoothed prior $p_2(\mathbf{m})$ and the geological prior $p_3(\mathbf{m})$, respectively. In each column, a random posterior sample, mean velocity, standard deviation and relative error maps are displayed from top to bottom row, respectively.}
	\label{fig:fwi_mean_std}
\end{figure}

Figures \ref{fig:fwi_mean_std}a -- \ref{fig:fwi_mean_std}c display the obtained inversion results. Each figure includes a random posterior sample, the mean velocity, standard deviation and the relative error maps of the posterior pdf from top to bottom row. Note that Figures \ref{fig:test_vpr_mean}b and \ref{fig:fwi_mean_std}b represent the same results obtained using VPR. In a previous study \citep{zhao2024physically}, we compared the inversion results obtained without using prior replacement displayed in Figure \ref{fig:fwi_mean_std}a with two entirely independent variational methods using exactly the same uniform prior distribution and observed data and obtained highly consistent results, proving that we obtain approximately correct posterior uncertainty statistics for this specific prior. In this study we focus on the inversion results obtained using different priors. 

The posterior random sample in Figure \ref{fig:fwi_mean_std}a shows significant velocity `speckle' - strong, short wavelength contrasts - since no correlation is introduced by the uniform prior distribution. The relatively non-informative prior information results in significantly higher uncertainties at greater depths, increasing up to around 800 m/s. The two posterior samples in Figures \ref{fig:fwi_mean_std}b and \ref{fig:fwi_mean_std}c are smoother since extra (smooth) prior information is injected into the two inversion results which precludes sharp velocity variations between neighbouring cells. The smoothed prior pdf imposes spatial smoothness explicitly, and the geological prior pdf injects similar information implicitly, as illustrated by the positive correlations displayed in Figure \ref{fig:prior_cor_geological}. Therefore, the true velocity structures are better resolved since they are indeed laterally fairly smooth.

The three mean velocity maps are quite smooth and similar to each other, generally resembling the true velocity map. The methods fail to recover some thin layers in the deeper part of the model due to the limited frequency band of the waveform data (10 Hz dominant frequency). Also, the mean can sometimes have a very low, even zero, probability density, especially in the case of the uniform prior distribution where such a smoothed mean model might be precluded by the data: this is because purely observed waveform data would inject negative correlations between neighbouring cells as displayed below in Figure \ref{fig:fwi_correlation}a and in \cite{gebraad2020bayesian, zhang20233, zhao2024physically}. Therefore, smoothed velocity structures (such as the mean model) which imply positive correlations between adjacent cells, might have low probability values.

Standard deviation values displayed in Figures \ref{fig:fwi_mean_std}b and \ref{fig:fwi_mean_std}c are smaller than those in Figure \ref{fig:fwi_mean_std}a (note that different colorbars are used in the former figures). This makes sense because by imposing prior information that velocity structures should be relatively smooth, we have removed the possibility of including large velocity contrasts between laterally proximal cells. In fact, the introduction of prior information, if it correctly reflects the true state of nature, should lead to models that better reflect the true state of nature overall (other than in pathalogical cases). In our case, the geological prior information introduced is derived from pictures that represent real geology and is therefore reasonably reflective of the true model, so in this case at least, the introduction of prior information should improve the result. Nevertheless, in some cases the uncertainty reduction displayed in Figures \ref{fig:fwi_mean_std}b and \ref{fig:fwi_mean_std}c might not be a good outcome. Normally there is less information at greater depths from waveform data recorded at the surface. The decreased uncertainties at depth occur because information from data and prior knowledge are combined. This only leads to more accurate models if the prior information is also accurate. In the case of our smoothed prior, smoothing is applied equally in vertical and horizontal directions which does not correctly reflect the structure of the true model. In that case we therefore would not expect that results are more accurate after applying this prior information (but they may be, for example if the relative errors in vertical smoothing are more than compensated by increased accuracy in horizontal smoothing). In addition, the posterior uncertainties in Figure \ref{fig:fwi_mean_std}c are higher than those in Figure \ref{fig:fwi_mean_std}b, possibly due to the different magnitude of smoothness applied to these two results (one controlled by the predefined parameter $\Sigma_{\mathbf{Sm}}$ in equation \ref{eq:sm} and the other by correlations calculated from the training images such as Figure \ref{fig:geological_images}). 

In addition to the magnitude of the standard deviation values, strong prior information also suppresses overall changes in the uncertainty structures: obvious spatial (vertical) variations are observed in the standard deviation map in Figure \ref{fig:fwi_mean_std}a. For example, uncertainties increase at depth since data sensitivity decreases at depth and lower uncertainties are observed around some high velocity layers. However, those features are less significant in Figure \ref{fig:fwi_mean_std}c and almost invisible in Figure \ref{fig:fwi_mean_std}b. This is because the geological prior information imposes weaker vertical smoothness, as illustrated in Figures \ref{fig:prior_cor_geological} and \ref{fig:prior_samples}c, whereas the smoothed prior distribution imposes the same magnitude of smoothness in both vertical and horizontal directions. As a result, we observe larger relative errors in Figures \ref{fig:fwi_mean_std}b and \ref{fig:fwi_mean_std}c, especially at layer boundaries where higher uncertainties \textit{should} be expected \citep{galetti2015uncertainty}.

\begin{figure}
	\centering\includegraphics[width=\textwidth]{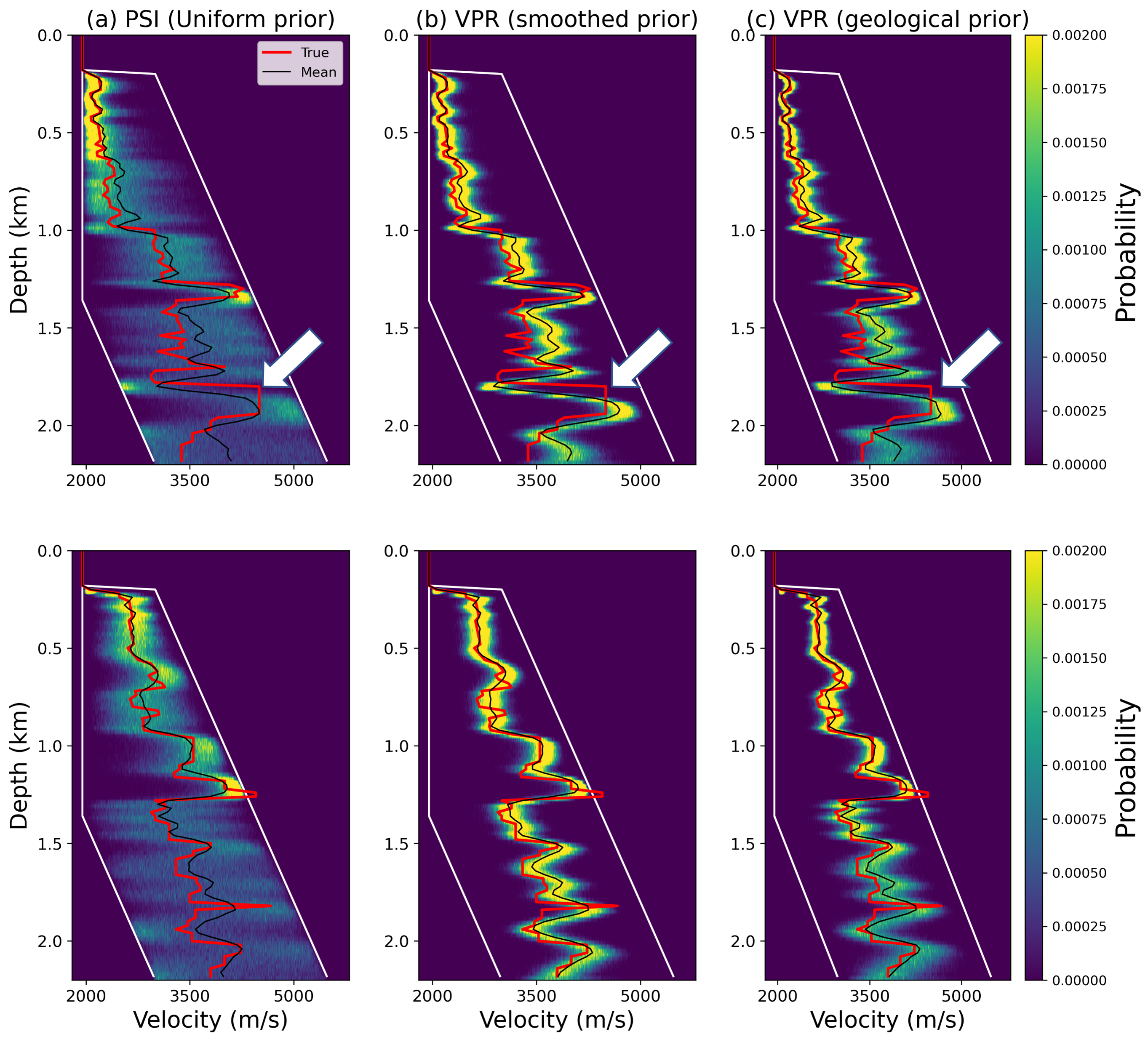}
	\caption{Posterior marginal distributions along two vertical velocity profiles at locations of 1 km (top row) and 2.6 km (bottom row) marked by two black dashed lines in Figure \ref{fig:fwi_110_250_vel_data_prior}a. Columns (a) -- (c) correspond to the same inversion results as those displayed in Figures \ref{fig:fwi_mean_std}a -- \ref{fig:fwi_mean_std}c.}
	\label{fig:fwi_marginal}
\end{figure}

Figure \ref{fig:fwi_marginal} displays the posterior marginal pdfs of the three inversion results at the same two locations as in Figure \ref{fig:test_vpr_marginal}. Similarly to the standard deviation maps, the marginal pdfs in Figures \ref{fig:fwi_marginal}b and \ref{fig:fwi_marginal}c are narrower than those in Figure \ref{fig:fwi_marginal}a due to the additional prior information injected. Figure \ref{fig:fwi_marginal}c presents larger vertical variations than Figure \ref{fig:fwi_marginal}b. As marked by the three white arrows in the first row, the (old) posterior pdf using the uniform prior distribution fails to find the true solution, thus neither do the two VPR results. This is reasonable considering the variational prior replacement methodology: the method can only \textit{replace} prior information imposed previously into the inversion results, but it cannot \textit{correct (improve)} the old estimate of the posterior distribution in cases where the old estimate is poor. Nevertheless, in places where the old posterior distribution includes the true model solution, VPR injects new prior information properly as displayed in the bottom row.

To analyse posterior correlation information in both horizontal and vertical directions, we calculate the correlation matrices for velocity values selected from 10 horizontally contiguous and 10 vertically contiguous grid cells (marked by top and left boundaries of the black box in Figure \ref{fig:fwi_110_250_vel_data_prior}a). The corresponding results are displayed in the top and bottom rows of Figure \ref{fig:fwi_correlation}. Similarly to Figures \ref{fig:fwi_mean_std} and \ref{fig:fwi_marginal}, each column (Figures \ref{fig:fwi_correlation}a -- \ref{fig:fwi_correlation}c) represents the calculated correlation matrix from one specific prior pdf. Since no correlation information is introduced by the uniform prior distribution, the posterior correlations displayed in Figure \ref{fig:fwi_correlation}a are purely determined by observed data. Negative correlations are observed between neighbouring cells and positive correlations are vaguely presented between every second neighbouring cells. Since FWI is a highly underdetermined inverse problem (the effective number of independent data points is significantly smaller than the number of unknown model parameters), velocity values oscillate between adjacent cells to achieve a better data fit, especially for grid cells within one wavelength. Similar posterior correlation patterns using a uniform prior distribution were observed in previous studies \citep{gebraad2020bayesian, zhang20233, zhao2024physically}. This would also happen in linearised (deterministic) FWI if no regularisation term (prior information) was added. 

As discussed previously, the smoothed prior distribution imposes the same magnitude of smoothness in both directions to prevent sharp velocity changes (i.e., positive correlations between adjacent cells). The posterior correlations presented in Figure \ref{fig:fwi_correlation}b are thus a result of the combination of correlation information from both the prior pdf (positive correlations) and observed data (negative correlations). Positive correlation values between neighbouring cells in Figure \ref{fig:fwi_correlation}b indicate that correlation information from the smoothed prior is stronger than that from the waveform data. The geological prior pdf injects different levels of smoothness in horizontal and vertical directions, as illustrated in Figure \ref{fig:prior_cor_geological}. In the horizontal direction, the magnitude of the smoothness injected by the prior pdf is presumably stronger than the magnitude of anti-correlation injected by the data, resulting in positive correlations between adjacent cells (top row in Figure \ref{fig:fwi_correlation}c). On the other hand, we observe almost zero correlations between vertically neighbouring cells (bottom row in Figure \ref{fig:fwi_correlation}c), implying that the vertical correlation information injected from the prior and data have similar strength, thus being cancelled out completely.

\begin{figure}
	\centering\includegraphics[width=\textwidth]{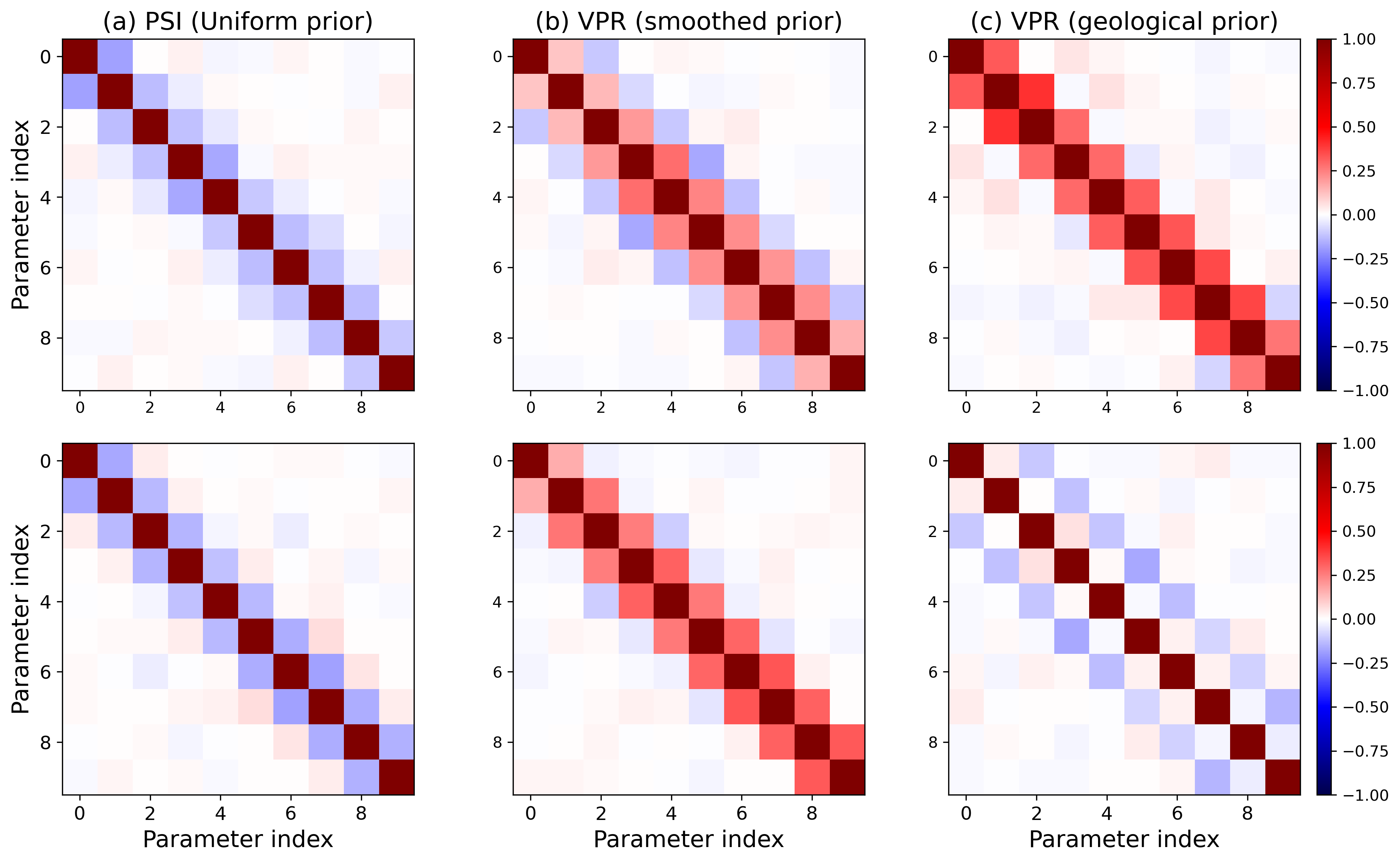}
	\caption{Posterior correlations for velocity values in 10 horizontally (top row) and vertically (bottom row) contiguous cells along the top and left of the black box in Figure \ref{fig:fwi_110_250_vel_data_prior}a, resulting in correlation matrices with sizes of 10 $\times$ 10. From left to right, (a) -- (c) correspond to the same inversion results as those displayed in Figures \ref{fig:fwi_mean_std} and \ref{fig:fwi_marginal}.}
	\label{fig:fwi_correlation}
\end{figure}

\subsection{Computational cost}
In Table \ref{table:fwi} we summarise the detailed computational cost for the four different inversion tests performed in the previous two sections. The first two rows in Table \ref{table:fwi} present the computational cost for the prior specific inversion (PSI) method using the uniform and smoothed prior distributions. In each test, we update the variational parameters for 5,000 iterations with 2 samples per iteration, resulting in a total number of 10,000 forward and adjoint (FWI) simulations. This process is performed using 36 CPU cores with a wall clock time of approximately 2 days. The other two results are obtained using variational prior replacement (VPR) with smoothed and geological prior distributions, based on the inversion results from the uniform prior pdf. To solve these two VPR problems, we use 5,000 iterations and 10 samples per iteration to minimise the KL divergence in equation \ref{eq:kl_vpr}. This does not require any FWI simulations to calculate the likelihood (data misfit) value for any sample. We need only to evaluate the probability value of the old posterior pdf represented by $q_{old}(\mathbf{m})$. To achieve this, in VPR we construct $q_{old}(\mathbf{m})$ using a parametric variational inference method (in this case we use PSVI introduced in Section \ref{section:psvi}). Note that the computational cost for evaluating the probability value $q_{old}(\mathbf{m})$ is almost zero compared to the forward and gradient simulations in FWI. Therefore, the two VPR results can be obtained within 5 minutes using 1 CPU core, which can be performed efficiently even on a laptop. The costs for both PSI and VPR depend on subjective assessments of the point of convergence, so the absolute computational time listed in Table \ref{table:fwi} might not be entirely accurate. Nevertheless, it is still obvious that VPR is significantly cheaper than PSI since no further FWI simulation are involved once we have obtained the old posterior distribution. In Section \ref{section:verify_vpr}, we showed that PSI and VPR provide almost identical results. This makes the proposed method attractive when multiple different priors are available or need to be tested using the same observed data, as presented in this paper.

\begin{table}
	\caption{A comparison of computational cost of the different tests performed in this study. PSI and VPR stand for \textit{prior specific inversion} and \textit{variational prior replacement}, respectively. }
	\centering
	\begin{tabular}{ccccccc}
		\hline
		Prior pdf & Method&\begin{tabular}{@{}c@{}}Number\\of iterations\end{tabular} & \begin{tabular}{@{}c@{}}Samples\\per iteration\end{tabular} & \begin{tabular}{@{}c@{}} Number of \\ FWI simulations \end{tabular}&CPU cores&Elapsed time \\
		\hline 
		Uniform & PSI & 5000 & 2 & 10,000 & 36 & 2 days \\
		Smoothed & PSI & 5000 & 2 & 10,000 & 36 & 2 days \\
		Smoothed & VPR & 5000 & 10 & 0 & 1 & 5 minutes \\
		Geological & VPR & 5000 & 10 & 0 & 1 & 5 minutes \\
		\hline
	\end{tabular}
	\label{table:fwi}
\end{table}

\section{Discussion}
We demonstrated that variational prior replacement (VPR) can change prior information efficiently post Bayesian inference. The updated posterior distribution is found by solving a variational problem, in which a variational distribution $q_{new}(\mathbf{m})$ is introduced and optimised iteratively to approximate $p_{new}(\mathbf{m}|\mathbf{d}_{obs})$, as expressed in equation \ref{eq:bayes_vpr2}. Therefore, we do not expect VPR and prior specific inversion (PSI) to provide exactly the same results, especially if PSI itself is performed using variational inference as in this study which also results in an approximation. This is part of the reason why we observe some small discrepancies between the results obtained using VPR and PSI displayed in Figures \ref{fig:test_vpr_mean} -- \ref{fig:test_vpr_correlation}. 

A similar effect was observed in the original prior replacement paper \citep{walker2014varying} which used mixture density networks \cite[MND --][]{bishop1994mixture} to estimate the old posterior pdf, and employed (semi-)analytic methods to calculate the new posterior pdf using equation \ref{eq:bayes_new}. In that case, the obtained pdf was still not the actual posterior distribution $p_{new}(\mathbf{m}|\mathbf{d}_{obs})$ given observed data and new prior information, since the old posterior distribution $p_{old}(\mathbf{m}|\mathbf{d}_{obs})$ used in equation \ref{eq:bayes_new} remained an approximation represented by the MND. This explains why the results obtained using prior replacement and direct Monte Carlo sampling (i.e., prior specific inversion) displayed in \cite{walker2014varying} (Figures 2 and 3) are not identical. Nevertheless, most of the posterior statistics in this current paper are nearly identical between PSI and VPR, implying that VPR is effective at updating prior information.

In addition, (semi-)analytic calculation of equation \ref{eq:bayes_new} requires the evaluation of the normalisation constant $k$ by integrating $p_{new}(\mathbf{m})$, $p_{old}(\mathbf{m})$ and $p_{old}(\mathbf{m}|\mathbf{d}_{obs})$ over the entire parameter space analytically, which might be intractable for high dimensional inference problems, and indeed only under certain circumstances can this be done \citep{walker2014varying}. In the proposed VPR framework, we introduce a second variational distribution $q_{new}(\mathbf{m})$, which is found by minimising the KL-divergence in equation \ref{eq:kl_vpr}; the specific value of $k$ then need not be estimated explicitly, so VPR can be implemented in a more straightforward manner. On the other hand, this implies that VPR is itself an approximate method which uses $q_{new}(\mathbf{m})$ to approximate $p_{new}(\mathbf{m}|\mathbf{d}_{obs})$.

One essential requirement for prior replacement developed here and in \cite{walker2014varying} is that the probability value for the old posterior pdf must be able to be evaluated cheaply (otherwise, there is no reason to use VPR instead of prior specific inversion). In this study, we use physically structured variational inference for this purpose \citep{zhao2024physically}, which constructs a transformed Gaussian distribution with a specific correlation structure to approximate the old posterior distribution $p_{old}(\mathbf{m}|\mathbf{d}_{obs})$ so that the probability value $q_{old}(\mathbf{m})$ can be calculated efficiently. Other well-established parametric variational inference methods, such as normalising flows \citep{rezende2015variational, dinh2014nice, kobyzev2019normalizing, papamakarios2019normalizing, zhao2022interrogating, siahkoohi2020faster, levy2022variational}, automatic differentiation variational inference \citep{kucukelbir2017automatic, zhang2019seismic, bates2022probabilistic, sun2023new} and boosting variational inference \citep{guo2016boosting, miller2017variational, locatello2018boosting2, zhao2024bayesian}, can also be used to construct $q_{old}(\mathbf{m})$. The choice of method should be based on the specific problem at hand, since the No Free Lunch theorem states that no method is better than any other when averaged across all problems \citep{wolpert1997no}. 
 
Note that the prior replacement step (the second step described by equations \ref{eq:bayes_new} and \ref{eq:bayes_new_q}) does not necessarily need to be solved using parametric variational methods or even variational inference. Various Monte Carlo sampling methods can also be used for this purpose as long as the dimensionality of the inverse problem is not too high \citep{curtis2001prior}.

\cite{walker2014varying} used mixture density networks (MDN) to approximate the old posterior distribution, and it has been shown to be difficult to capture posterior correlations between different parameters using this method \citep{zhang2021bayesian, bloem2022introducing}. Nevertheless, as shown in numerous studies \citep{devilee1999efficient, meier2007fully, meier2007global, shahraeeni2011fast, shahraeeni2012fast, kaufl2014framework, kaufl2016solving, earp2019probabilistic, earp2019probabilistic2, cao2020near, lubo2021exhaustive, hansen2022use, bloem2022introducing}, an advantage of using MDN is that they can determine the posterior pdf corresponding to any dataset extremely rapidly once the networks have been trained. In other words, varying observed data in Bayesian inference can be accomplished using MDN with almost no additional cost. Prior to the work of \cite{walker2014varying} and this current work, if we wished to change prior information we would have to re-train the MDN which typically requires millions of training samples and the calculation of their forward function values. A possible extension of the current work might combine VPR and MDN, in which MDN is used to calculate the old posterior distribution (using a set of prior samples obtained from the old prior pdf) and VPR is employed to evaluate any potential new posterior distribution when prior information changes. Under this framework, both prior information and observed data can be replaced efficiently with one single training of an MDN (using the old prior). This opens the possibility that advanced real-time monitoring of subsurface changes and the corresponding uncertainties can be implemented efficiently for some problems.

\cite{papamakarios2016fast} introduced a similar approach compared to VPR for likelihood-free inference using MDN. Traditionally one would use a large number (millions) of prior samples and their forward function values to train an MDN. If we are only interested in a posterior pdf for one specific dataset, such a strategy is inefficient since most of the prior samples and their forward evaluations would result in near-zero probabilities in that specific posterior pdf. Therefore, \cite{papamakarios2016fast} defined a proposal prior distribution $\tilde{p}(\mathbf{m})$, from which samples are drawn to train an MDN. The proposal prior distribution is updated iteratively to generate samples that are highly informative while training an MDN for a specific observation (for example, if $\tilde{p}(\mathbf{m})$ could be set equal to the true posterior pdf then all samples would be useful). After the training process, the proposal prior is replaced by the original prior pdf to obtain the true posterior pdf -- as described in equations \ref{eq:bayes_new} and \ref{eq:bayes_vpr2}. However, in that case the corresponding posterior pdf needs to be normalised \citep{walker2014varying}, and for cases in which the proposal prior is narrower than the original prior distribution replacing it would lead to a numerical issue of dividing by zero; both issues are likely to be especially prevalent in high dimensional problems and are mitigated in our methodology.

Based on a Markov chain Monte Carlo (McMC) framework, \cite{mosegaard1995monte} introduced an approach to draw samples from one probability distribution when one only has samples from another distribution. They achieved this by resampling using a simple Metropolis accept-reject criterion based on the ratio of two probability values $p_{new}(\mathbf{m})/p_{old}(\mathbf{m})$, potentially avoiding further likelihood function evaluations. In this sense, that method can be interpreted as a Monte Carlo version of prior replacement that provides a sampling-based solution, compared to the VPR approach which is based on variational inference. Both approaches of \cite{mosegaard1995monte} and in this paper have a similar pre-requisite: that the support of the old prior distribution $p_{old}(\mathbf{m})$ should include that of the new prior so that numerical instability issues are avoided when performing the division $p_{new}(\mathbf{m})/p_{old}(\mathbf{m})$.

In our numerical examples, geological prior information is represented by a Gaussian distribution with a local correlation structure estimated from images of real geology. A direct generalisation is to use a mixture of Gaussian distributions to model the geological prior distribution. In addition, normalising flows \citep{dinh2014nice, dinh2017density, papamakarios2017masked, kingma2018glow} are often used as a deep generative model in the machine learning community, which construct a complex probability distribution by passing a simple and analytically known probability distribution (such as a uniform or a standard normal distribution) through a series of invertible and differentiable transforms. After training a normalising flows model, the prior probability value of any model sample can be evaluated and used in the VPR framework. Future work might explore the use of these methods to build a more sophisticated prior distribution. On the other hand, geological prior information might be simulated through geological processing models \citep{tetzlaff1989simulating, paola2000quantitative, burgess2001numerical, hill2009modeling, tetzlaff2023stratigraphic}, which can then be parametrised by advanced neural network models and used during Bayesian inference \citep{laloy2018training, mosser2020stochastic, levy2022variational, scheiter2022upscaling, hillier2023geoinr, liu2024geostatistical, bloem2022introducing, bloem2024bayesian}. These approaches could lead to more accurate and realistic representations of geological structures and their associated uncertainties.

The main purpose of VPR is to update (replace) prior information efficiently in Bayesian inference. As demonstrated herein, new inversion results can be obtained with no further forward simulation. On the other hand, this also indicates that we cannot obtain new information about data misfit values purely from VPR results - indeed, that is the point of the method. In future, VPR might be used to compare different prior hypotheses (e.g., including different magnitudes of smoothness for a smoothed prior distribution): say we have obtained approximate posterior pdfs for a set of different prior assumptions by applying VPR. For the different inversion results, we could perform a small number of additional forward simulations using posterior samples drawn from VPR results, based on which the different prior hypotheses can be compared, and one or two close-to-optimal options could be selected. Although this procedure does require additional forward simulations, it still provides a far more efficient approach to test different prior options compared to carrying out a sequence of independent inversions as is typically done using linearised inversion. In addition, for cases when VPR becomes less accurate (e.g., when the dimensionality and complexity of the inverse problem increase such as in 3D FWI problems), one might use a relatively lower cost forward function with different data types to refine (fine tune) the outcomes obtained from VPR, by invoking Bayes' rule again.


\section{Conclusions}
We develop a variational prior replacement (VPR) methodology designed to efficiently update prior information in Bayesian inference solutions. This approach involves replacing the existing prior information in a posterior distribution obtained from a previous inference process with a new prior distribution. The new posterior distribution is then found using variational inference. VPR eliminates the need to re-solve Bayesian inverse problems from scratch each time prior information changes. The results from a 2D full waveform inversion example support the effectiveness of VPR for varying prior information, in which VPR provides consistent statistics of the posterior probability distribution compared to those obtained using the conventional prior specific inversion scheme. This similarity holds for individual posterior samples, first- and second-order statistics, as well as simulated waveform data. The key advantage of VPR lies in its computational efficiency: achieving the same results in a matter of minutes compared to two days required by the conventional approach. Additionally, we show that VPR can be used to investigate the impact of different prior distributions on Bayesian inference results. This methodology has significant potential for applications in more computationally demanding inverse problems, such as 3D Bayesian FWI, especially when multiple priors are available or need to be tested and discriminated for the same dataset.

\section{Data Availability}
The code underlying this article will be shared on reasonable request to the corresponding author.

\begin{acknowledgments}
We thank the Edinburgh Imaging Project (EIP - \url{https://blogs.ed.ac.uk/imaging/}) sponsors (BP and TotalEnergies) for supporting this research. For the purpose of open access, we have applied a Creative Commons Attribution (CC BY) licence to any Author Accepted Manuscript version arising from this submission.
\end{acknowledgments}

\bibliographystyle{gji}
\bibliography{reference}

\appendix

\section{Verifying VPR using a Normal Initial prior Distribution}
\label{ap:verify_vpr}

\begin{figure}
	\centering\includegraphics[width=\textwidth]{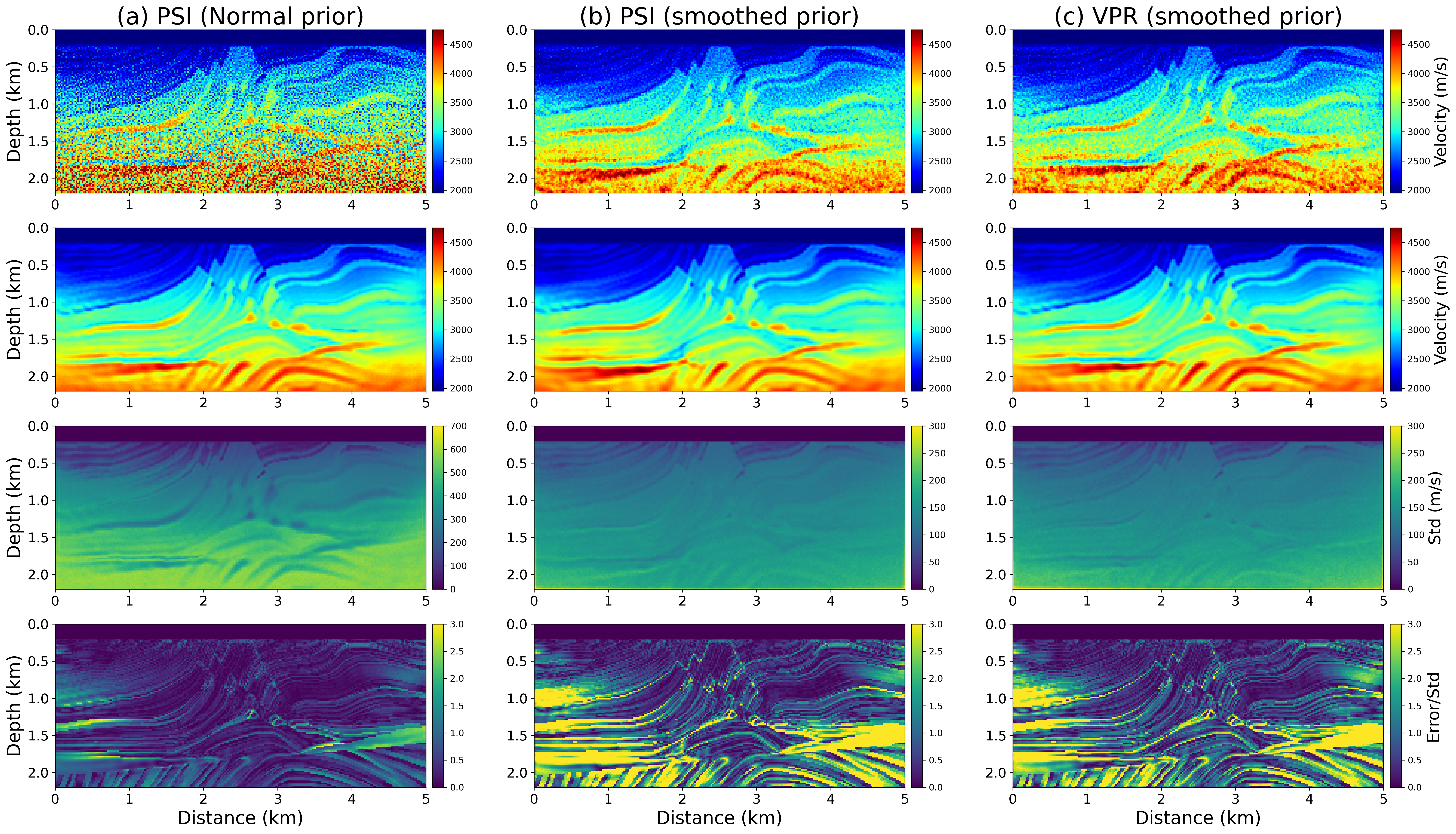}
	\caption{(a) Prior specific inversion (PSI) results obtained using a diagonal Gaussian prior distribution defined in this Appendix. (b) PSI results obtained using a smoothed version of the normal prior distribution. (c) Variational prior replacement (VPR) results obtained by replacing the normal prior distribution by the smoothed prior. In each column, a random posterior sample, mean velocity, standard deviation and relative error maps are displayed from top to bottom row, respectively.}
	\label{fig:fwi_110_250_3diff_normal_priors_psvi_3mean_std_10hz}
\end{figure}

In this appendix, we present a second example to test the performance of the proposed variational prior replacement (VPR) methodology. We define a diagonal Gaussian distribution as the old prior pdf, with mean and standard deviation are calculated from the uniform prior distribution used in the main text. Figure \ref{fig:fwi_110_250_3diff_normal_priors_psvi_3mean_std_10hz}a shows the inversion results obtained using this normal prior distribution. From top to bottom, each panel represents a random posterior sample, the mean velocity, standard deviation and the relative error maps of the posterior distribution, where the relative error is calculated as the difference between the mean and true velocity models divided by the standard deviation at each point. Given this inversion result, we perform VPR by replacing a smoothed version of the normal distribution (which is defined using exactly the same way as that expressed in equation \ref{eq:smooth_prior}) by the original normal prior distribution, and display the corresponding results in Figure \ref{fig:fwi_110_250_3diff_normal_priors_psvi_3mean_std_10hz}c. Similarly to the main text, in Figure \ref{fig:fwi_110_250_3diff_normal_priors_psvi_3mean_std_10hz}b we display prior specific inversion (PSI) results obtained using this same smoothed prior distribution. Since Figures \ref{fig:fwi_110_250_3diff_normal_priors_psvi_3mean_std_10hz}b and \ref{fig:fwi_110_250_3diff_normal_priors_psvi_3mean_std_10hz}c present highly consistent results, we conclude that VPR is accurate and effective in this example.

\label{lastpage}
\end{document}